\documentclass[preprint,prd,aps,showpacs,showkeys,nofootinbib]{revtex4} 
\usepackage{graphicx}
\usepackage{dcolumn}
\usepackage{bm}

\begin{document}

\preprint{hep-ph/0410111}

\title{Effective Lagrangian of the $/\!\!\!\!R$MSSM for neutrino mass generation}
 
\author{Tai-Fu Feng}

\affiliation{Department of Physics, 40014 University of Jyv\"askyl\"a,Finland}

\author{Jukka Maalampi}
\affiliation{Department of Physics, 40014 University of Jyv\"askyl\"a,Finland}

\date{\today}

\begin{abstract}

We derive the lepton number violating dimension-five and dimension-seven operators, 
relevant for neutrino mass generation,  in the Minimal Supersymmetry Standard Model 
without $R$-parity (the $/\!\!\!\!R$MSSM) by using the effective Lagrangian method. 
We keep all the possible
$CP$ violating phases, and we establish the general relationship between the
high-scale parameters and the low-energy observals associated with the Standard 
Model particles. We study in a specific model the dependence of neutrino masses  on 
the $CP$ phases.
\end{abstract}



\maketitle

\section{Introduction}
\indent\indent

Neutrino physics measurements show that neutrinos change their flavour in 
oscillationary manner \cite{exp}, implying that neutrinos have a mass. In the 
Standard Model (SM) of particle interactions neutrinos are strictly massless due to 
lepton number conservation and non-existence of right-handed neutrinos. The same is 
true for the most popular extension of the SM, the Minimal Supersymmetric Standard 
Model (MSSM) \cite{mssm}, where the lepton number conservation follows from the 
assumed R-parity symmetry. The R-parity of a field is defined as 
$R=(-1)^{3B+L+2s}=(-1)^{3(B-L)+2s}$, where $B,\;L$ and $\;s$ are the baryon number, 
lepton number and spin, respectively \cite{Rp}. The conservation of R-parity is 
imposed in order to prohibit the proton from fast decay, which may happen if both 
$B$ and $L$ are broken. If one abandons the R-symmetry by allowing lepton number to 
be violated, but leaving baryon number unbroken, the MSSM can accommodate 
$\vert\Delta L\vert=2$ Majorana masses for neutrinos \cite{rpv}.

The superpotential of the MSSM with broken R-parity and conserved baryon number 
(acronymed as $/\!\!\!\!R{\rm MSSM}$) is in the most general case
of the following form: 
\begin{eqnarray}
&&W=W_{_{R_P}}+W_{_{/\!\!\!\!R_P}}
\label{eq1}
\end{eqnarray}
with
\begin{eqnarray}
&&W_{_{R}}=\epsilon_{_0}H_{_u}H_{_d}+h_{_{IJ}}^eH_{_d}
L_{_I}E_{_J}^c+h_{_{IJ}}^dH_{_d}Q_{_I}D_{_J}^c
\nonumber\\
&&\hspace{1.2cm}
+h_{_{IJ}}^uH_{_u}Q_{_I}U_{_J}^c
\;,\nonumber\\
&&W_{_{/\!\!\!\!R}}=\epsilon_{_I}H_{_u}L_{_I}+{1\over2}\lambda_{_{IJK}}
L_{_I}L_{_J}E_{_K}^c
\nonumber\\
&&\hspace{1.2cm}
+\lambda_{_{IJK}}^\prime L_{_I}Q_{_J}D_{_K}^c\;.
\label{eq2}
\end{eqnarray}
The part $W_{_{R}}$ is the superpotential of the standard R-conserving MSSM, 
whereas $W_{_{/\!\!\!\!R}}$ consists of renormalizable bilinear and trilinear 
couplings that violate the lepton number conservation and thereby the R-parity.
In (\ref{eq2}) $H_{_u},\;H_{_d}$ are Higgs doublet superfields and 
$Q,\;L$ are left-handed quark and lepton superfields, all transforming as $SU(2)$
doublets, and $U,\;D,\;E$ are right-handed quark and lepton superfields
transforming as $SU(2)$ singlets. The flavour of the lepton and quark fields is 
denoted by indices $I,J,K$, and the summation
over gauge indices is implicit. The R-parity
violating three-lepton coupling obeys
$\lambda_{_{IJK}}=-\lambda_{_{JIK}}$. 

The breaking of supersymmetry is due to interaction and mass terms of component 
fields. In addition to the soft terms present in the MSSM, one should now include 
also the terms that break lepton number. The most general form of the soft-term 
Lagrangian in our case is then
\begin{eqnarray}
&&{{\cal L}_{_S}}={{\cal L}_{_S}^R}+{{\cal L}_{_S}^{/\!\!\!\!R}},
\label{eq3}
\end{eqnarray} 
where
\begin{eqnarray}
&&-{\cal L}_{_S}^R=(m_{_{\tilde Q}}^2)_{_{IJ}}\tilde{Q}_{_I}^\dagger\tilde{Q}_{_J}
+(m_{_{\tilde U}}^2)_{_{IJ}}\tilde{U}_{_I}^{c\dagger}\tilde{U}_{_J}^c
\nonumber\\
&&\hspace{1.2cm}
+(m_{_{\tilde D}}^2)_{_{IJ}}\tilde{D}_{_I}^{c\dagger}\tilde{D}_{_J}^c
+(m_{_{\tilde L}}^2)_{_{IJ}}\tilde{L}_{_I}^{\dagger}\tilde{L}_{_J}
\nonumber\\
&&\hspace{1.2cm}
+(m_{_{\tilde E}}^2)_{_{IJ}}\tilde{E}_{_I}^{c\dagger}\tilde{E}_{_J}^c
+\Big[A_{_{IJ}}^eH_{_d}\tilde{L}_{_I}\tilde{E}_{_J}^c
\nonumber\\
&&\hspace{1.2cm}
+A_{_{IJ}}^dH_{_d}\tilde{Q}_{_I}\tilde{D}_{_J}^c
+A_{_{IJ}}^uH_{_u}\tilde{Q}_{_I}\tilde{U}_{_J}^c+h.c.\Big]
\nonumber\\
&&\hspace{1.2cm}
+m_{_{H_u}}^2H_{_u}^\dagger H_{_u}+m_{_{H_d}}^2H_{_d}^\dagger H_{_d}
\nonumber\\
&&\hspace{1.2cm}
+\Big(B_{_0}H_{_u}H_{_d}+h.c.\Big)
+\Big[{1\over2}m_{_1}\lambda_{_B}\lambda_{_B}
\nonumber\\
&&\hspace{1.2cm}
+{1\over2}m_{_2}\lambda_{_{A^\alpha}}\lambda_{_{A^\alpha}}+{1\over2}m_{_3
}
\lambda_{_{G^a}}\lambda_{_{G^a}}+h.c.\Big]
\label{eq3a}
\end{eqnarray}
and
\begin{eqnarray}
&&-{\cal L}_{_S}^{/\!\!\!\!R}=
+{1\over2}A_{_{IJK}}\tilde{L}_{_I}\tilde{L}_{_J}\tilde{E}_{_K}^c
+A_{_{IJK}}^\prime\tilde{L}_{_I}\tilde{Q}_{_J}\tilde{D}_{_K}^c
\nonumber\\
&&\hspace{1.2cm}
+{\bf B}_{_I}H_{_u}\tilde{L}_{_I}+({\bf m}_{_{\tilde L}}^2)_{_{0I}}H_{_d}^\dagger
\tilde{L}_{_I}+h.c.\;.
\label{eq3b}
\end{eqnarray}
Here $\lambda_{_{G^a}}\;(a=1,\;2,\;\cdots,\;8),\;\lambda_{_{A^\alpha}}
\;(\alpha=1,\;2,\;3),\;\lambda_{_B}$ denote the gaugino fields corresponding to the 
gauge symmetry $SU(3)\times SU(2)\times 
U(1)$ in an obvious manner.

In this paper we shall study neutrino masses in the framework of the 
$/\!\!\!\!R{\rm MSSM}$. The induction of neutrino masses by
R-parity violation couplings have been extensively analyzed in the 
literature \cite{feng1}. Nevertheless, in most of these earlier studies the 
possible $CP$-phase effects on the neutrino masses have been ignored. In our 
analysis, we will keep all relevant $CP$ violation
phases of the Lagrangian as free parameters, and we shall show that they do affect 
the neutrino masses.

We will assume that the new physics, i.e. the physics that goes beyond the SM, is 
associated with an energy scale that is considerably larger than the electroweak 
scale of the SM, so that the new physics effects can be reliably analyzed by using 
the effective theory method. In that method, one integrates heavy fields out and 
keeps only the SM particles in the resulting effective theory \cite{Burgess}. The 
effects of the heavy fields are parameterized in terms of high-dimensional 
operators of the
light degrees of freedom. Due to the lepton number violating terms in the 
full Lagrangian, there will appear dimension-odd operators in the 
effective theory \cite{Weinberg}. After the electroweak symmetry breaking, these 
operators will induce a small (Majorana) mass for neutrinos, in accordance with 
experimental results \cite{Broncano}.

The paper is organized as follows. In Section \ref{sec2} we will present 
preliminary considerations concerning Higgs field and fermion field doublets and 
define the SM Higgs doublet in terms of the original doublets and their vacuum 
expectation values.
In Section \ref{sec3} we derive, following the analysis of \cite{feng2},  
the Wilson coefficients at the matching scale for the
operators relevant for neutrino masses by integrating out the supersymmetric 
particles and the heavy
Higgs doublet fields. The connection between 
the high-energy parameters
and low-energy observables are also discussed in this Section.
The results of numerical analysis and conclusions are presented
in the last Section.

\section{Preliminaries\label{sec2}}
\indent\indent
In the MSSM with no conserved lepton number there
is no a priori distinction between the down-type Higgs boson and  slepton doublets, 
which are assigned
with identical quantum numbers. One can therefore freely rotate the
weak eigenstate basis $(H_d,{\tilde L}_I)$ ($I=1,2,3$) by an $SU(4)$ 
transformation. The lepton number violating couplings depend on the basis one 
chooses. Nevertheless, the couplings of the SM Higgs field that appear in the 
low-energy theory
should not depend on an specific choice of the basis. 

The Higgs boson and slepton  doublets can be presented as follows:
\begin{eqnarray}
&&H_{_u}=\left(\begin{array}{c}H_{_u}^+\\  \\{1\over\sqrt{2}}\Big(\upsilon_{_u}
+H_{_u}^0+iA_{_u}^0\Big)\end{array}\right)\;, \nonumber\\
&&H_{_d}=\left(\begin{array}{c}{1\over\sqrt{2}}\Big(\upsilon_{_0}
+H_{_d}^0+iA_{_d}^0\Big)\\ \\H_{_d}^-\end{array}\right)\;, \nonumber\\
&&{\tilde L}_{_I}=\left(\begin{array}{c}{1\over\sqrt{2}}\Big(\upsilon_{_I}
+H_{_I}^0+iA_{_I}^0\Big)\\ \\{\tilde L}_{_I}^-\end{array}\right)\;(I=1,\;2,\;3)\;.
\label{eq4}
\end{eqnarray}
Here $\upsilon_{_u},\;\upsilon_{0},\;\upsilon_{_I}$
denote the vacuum expected values (VEVs) of the neutral components of the doublets 
$H_u,\; H_d$ and ${\tilde L}_{_I}$, respectively. One is obviously free to rotate 
these doublets in such a way that only one doublet achieves  a non-vanishing vacuum 
expectation value.
Let us denote
\begin{eqnarray}
&&\upsilon_{_{d_0}}=\upsilon_{_0}\;,\nonumber\\
&&\upsilon_{_{d_1}}=\sqrt{\upsilon_{_0}^2+\upsilon_{_1}^2}\;,\nonumber\\
&&\upsilon_{_{d_2}}=\sqrt{\upsilon_{_0}^2
+\upsilon_{_1}^2+\upsilon_{_2}^2}\;,\nonumber\\
&&\upsilon_{_{d_3}}=\sqrt{\upsilon_{_0}^2
+\upsilon_{_1}^2+\upsilon_{_2}^2+\upsilon_{_3}^2}\equiv\upsilon_{_d}\;,
\label{eq5}
\end{eqnarray}
and replace the Higgs boson and slepton doublets  of Eq.~(\ref{eq4}) with the set 
$(\Phi$, $\Phi_{_{H_1}}$, $\Phi_{_{H_2}}$, $\Phi_{_{H_3}}$, $\Phi_{_{H_4}})$ defined as 
\begin{eqnarray}
\left(\begin{array}{l}\Phi\\\Phi_{_{H_1}}\\\Phi_{_{H_{I+1}}}\end{array}\right)=
{\cal Z}_{_H}^0\left(\begin{array}{c}H_{_u}\\(i\sigma^2)H_{_d}^*
\\(i\sigma^2)\tilde{L}_{_I}^*\end{array}\right)\;,
\label{eq6}
\end{eqnarray}
where ${\cal Z}_{_H}^0$ denotes the transformation matrix 
\begin{widetext}
\begin{eqnarray}
&&{\cal Z}_{_H}^0=\left(\begin{array}{ccccc}-s_{_\beta}
&c_{_\beta}{\upsilon_{_0}\over\upsilon_{_d}}&c_{_\beta}{\upsilon_{_1}
\over\upsilon_{_d}}&c_{_\beta}{\upsilon_{_2}\over\upsilon_{_d}}&
c_{_\beta}{\upsilon_{_3}\over\upsilon_{_d}}\\
c_{_\beta}&s_{_\beta}{\upsilon_{_0}\over\upsilon_{_d}}&s_{_\beta}{\upsilon
n_{_1}
\over\upsilon_{_d}}&s_{_\beta}{\upsilon_{_2}\over\upsilon_{_d}}&
s_{_\beta}{\upsilon_{_3}\over\upsilon_{_d}}\\
0&-{\upsilon_{_1}\over\upsilon_{_{d_1}}}&{\upsilon_{_0}\over\upsilon_{_{d
_1}}}&
0&0\\
0&-{\upsilon_{_0}\upsilon_{_2}\over\upsilon_{_{d_1}}\upsilon_{_{d_2}}}
&-{\upsilon_{_1}\upsilon_{_2}\over\upsilon_{_{d_1}}\upsilon_{_{d_2}}}
&{\upsilon_{_{d_1}}\over\upsilon_{_{d_2}}}&0\\
0&-{\upsilon_{_0}\upsilon_{_3}\over\upsilon_{_{d_2}}\upsilon_{_{d}}}&
-{\upsilon_{_1}\upsilon_{_3}\over\upsilon_{_{d_2}}\upsilon_{_{d}}}&
-{\upsilon_{_2}\upsilon_{_3}\over\upsilon_{_{d_2}}\upsilon_{_{d}}}&
{\upsilon_{_{d_2}}\over\upsilon_{_{d}}}
\end{array}\right)\;.
\label{eq7}
\end{eqnarray}
\end{widetext}
The doublets $\Phi,\Phi_{_{H_{\alpha}}},\;(\alpha=1,\;2,\;3,\;4)$ can be presented 
in the form 
\begin{eqnarray}
&&\Phi=\left(\begin{array}{c}G^+\\ \\{1\over\sqrt{2}}\Big(h+\upsilon+iG^0\Big)
\end{array}\right)\;,\nonumber\\
&&\Phi_{_{H_\alpha}}=\left(\begin{array}{c}H_\alpha^+\\ 
\\{1\over\sqrt{2}}\Big(H_\alpha^0
+iA_\alpha^0\Big)\end{array}\right).
\label{eq8}
\end{eqnarray}
The doublet $\Phi$ corresponds to the Higgs doublet of the SM, having a vacuum 
expectation value $\upsilon=\sqrt{\upsilon_{_u}^2+\upsilon_{_d}^2}=246\;{\rm GeV}$.
It remains massless prior to the electroweak symmetry breaking and does not mix
with the other doublets. The other doublets obtain a large mass originating in the 
soft supersymmetry breaking terms of the Lagrangian. Indeed, the masses of the 
doublets are given by
\begin{eqnarray}
\left(\begin{array}{ccc}\Phi&\Phi_{_{H_1}}^\dagger&\Phi_{_{H_{I+1}}}^\dagger
\end{array}\right)
{\bf M}_{_H}^2
\left(\begin{array}{c}\Phi\\\Phi_{_{H_1}}\\\Phi_{_{H_{J+1}}}\end{array}\right)\;,
\label{eq9}
\end{eqnarray}
where 
\begin{eqnarray}
&&{\bf M}_{_H}^2=\left(\begin{array}{ccc}0&0&0\\0&(B/s_{_\beta}c_{_\beta})
&(s_{_\beta}-c_{_\beta})E_{_0}E_{_I}^*\\ 0&(s_{_\beta}-c_{_\beta})E_{_0}^*E_{_J}& 
({\bf m}_{_H}^2)_{_{IJ}}+E_{_I}^*E_{_J}\end{array}\right).
\label{eq9-a}
\end{eqnarray}
We have used here the following notations:
\begin{eqnarray}
&&B=\sum\limits_{_{\rho=0}}^3{\upsilon_{_\rho}
\over\upsilon_{_d}}\Big({\bf B}\Big)_{_\rho}
\;,\nonumber\\
&&\Big({\bf m}_{_H}^2\Big)_{_{IJ}}={\upsilon_{_{d_{(I-1)}}}
\upsilon_{_{d_{(J-1)}}}\over\upsilon_{_{d_I}}\upsilon_{_{d_J}}}
\Big({\bf m}_{_{\tilde L}}^2\Big)_{_{IJ}}
\nonumber\\&&\hspace{1.2cm}
+{\upsilon_{_I}\upsilon_{_J}
\over\upsilon_{_{d_{(I-1)}}}\upsilon_{_{d_{(J-1)}}}\upsilon_{_{d_I}}
\upsilon_{_{d_J}}}\sum\limits_{\rho=0}^{I-1}\sum\limits_{\sigma=0}^{J-1}
\upsilon_{_\rho}\upsilon_{_\sigma}\Big({\bf m}_{_{\tilde L}}^2\Big)_{_{\rho\sigma}}
\nonumber\\&&\hspace{1.2cm}
-{\upsilon_{_{d_{(J-1)}}}\upsilon_{_I}\over\upsilon_{_{d_{(I-1)}}}
\upsilon_{_{d_I}}\upsilon_{_{d_J}}}\sum\limits_{\rho=0}^{I-1}
\Big({\bf m}_{_{\tilde L}}^2\Big)_{_{J\rho}}\upsilon_{_\rho}
\nonumber\\&&\hspace{1.2cm}
-{\upsilon_{_{d_{(I-1)}}}\upsilon_{_J}\over\upsilon_{_{d_{(J-1)}}}
\upsilon_{_{d_I}}\upsilon_{_{d_J}}}\sum\limits_{\rho=0}^{J-1}
\upsilon_{_\rho}\Big({\bf m}_{_{\tilde L}}^2\Big)_{_{\rho I}}\;,\nonumber\\
&&{\bf E}_{_0}=\sum\limits_{\rho=0}^3{\upsilon_{_\rho}\over\upsilon_{_d}}
\epsilon_{_\rho}\;,\nonumber\\
&&{\bf E}_{_I}={\upsilon_{_{d_{(I-1)}}}^2
\epsilon_{_I}-\upsilon_{_I}\sum\limits_{\rho=0}^{I-1}\upsilon_{_\rho}
\epsilon_{_\rho}\over\upsilon_{_{d_{(I-1)}}}\upsilon_{_{d_I}}}\;.
\label{eq10}
\end{eqnarray}
Here $I,\;J=1,\;2,\;3$ and $\rho,\sigma=0,\;1,\;2,\;3$, and we have denoted $({\bf 
m}_{_{\tilde L}}^2)_{00}=m_{H_d}^2$ and $({\bf m}_{_{\tilde L}}^2)^*_{I0}
=({\bf m}_{_{\tilde L}}^2)_{0I}$.
In the effective theory the heavy eigenstates of this mass Lagrangian are 
integrated out and only the fields in
the massless doublet $\Phi$ are considered as a physical degrees of freedom.

The bilinear R-parity violating terms of the Lagrangian (\ref{eq3b}) lead to mixing 
of the
down-type higgsino and the left-handed leptons, and some of these fields obtain a 
large mass. The mass eigenstates are obtained by the following  
transformation:
\begin{eqnarray}
\left(\begin{array}{c}\psi_{_{h_d}}\\\psi_{_{l_I}}\end{array}\right)
={\cal Z}_{_{\tilde h}}\left(\begin{array}{c}\psi_{_{H_d}}\\\psi_{_{L_I}}
\end{array}\right)\;(I=1,\;2,\;3)\;,
\label{eq11}
\end{eqnarray}
where the transformation matrix ${\cal Z}_{_{\tilde h}}$ is given by
\begin{widetext}
\begin{eqnarray}
&&{\cal Z}_{_{\tilde h}}=\left(\begin{array}{cccc}{|\epsilon_{_0}|\over
\mu_{_H}}&{|\epsilon_{_1}|\over\mu_{_H}}e^{i\varphi_{_1}}&
{|\epsilon_{_2}|\over\mu_{_H}}e^{i\varphi_{_2}}&{|\epsilon_{_3}|\over\mu_
{_H}}
e^{i\varphi_{_3}}\\
-{|\epsilon_{_1}|\over\mu_{_{H_1}}}e^{-i\varphi_{_1}}&{|\epsilon_{_0}|\over
\mu_{_{H_1}}}
&0&0\\
-{|\epsilon_{_0}\epsilon_{_2}|\over\mu_{_{H_1}}\mu_{_{H_2}}}e^{-i\varphi_
{_2}}
&-{|\epsilon_{_1}\epsilon_{_2}|\over\mu_{_{H_1}}\mu_{_{H_2}}}
e^{-i(\varphi_{_2}-\varphi_{_1})}&{\mu_{_{H_1}}\over\mu_{_{H_2}}}&0\\
-{|\epsilon_{_0}\epsilon_{_3}|\over\mu_{_{H_2}}\mu_{_{H}}}e^{-i\varphi_{_
3}}
&-{|\epsilon_{_1}\epsilon_{_3}|\over\mu_{_{H_2}}\mu_{_{H}}}
e^{-i(\varphi_{_3}-\varphi_{_1})}&-{|\epsilon_{_2}\epsilon_{_3}|\over\mu_
{_{H_2}}\mu_{_{H}}}
e^{-i(\varphi_{_3}-\varphi_{_2})}&{\mu_{_{H_2}}\over\mu_{_{H}}}
\end{array}\right)\;.
\label{eq12}
\end{eqnarray}
\end{widetext}
Here the following notation is used:
\begin{eqnarray}
&&\mu_{_{H}}=\sqrt{\sum\limits_{\rho=0}^3|\epsilon_{_\rho}|^2},\;
\mu_{_{H_0}}=|\epsilon_{_0}|,\nonumber\\
&&\mu_{_{H_1}}=\sqrt{|\epsilon_{_0}|^2
+|\epsilon_{_1}|^2},\nonumber\\
&&\mu_{_{H_2}}=\sqrt{|\epsilon_{_0}|^2+|\epsilon_{_1}|^2
+|\epsilon_{_3}|^2},\nonumber\\
&&\mu_{_{H_3}}=\mu_{_{H}}. 
\label{eq12-a}
\end{eqnarray}
The phases $\varphi_{_I}\;(I=1,\;2,\;3)$, appearing in the transformation matrix,
are defined as $\varphi_{_I}=arg(\epsilon_{_I})
-\theta_{_\mu}$, where $\theta_{_\mu}\equiv arg(\epsilon_{_0})$. The mass terms 
induced by the bilinear
terms are
\begin{eqnarray}
\mu_{_H}e^{i\theta_{_\mu}}\psi_{_{H_u}}\psi_{_{h_d}}+\mu_{_H}e^{-i\theta_{_\mu}}
\overline{\psi}_{_{H_u}}\overline{\psi}_{_{h_d}}
\label{eq13}
\end{eqnarray}
The phase $\theta_{_\mu}$ can be absorbed into the fields by the following 
redefinitions:
\begin{eqnarray}
\psi_{_{H_u}}\rightarrow e^{-{i\over2}\theta_{_\mu}}\psi_{_{H_u}},\;
\psi_{_{h_d}}\rightarrow e^{-{i\over2}\theta_{_\mu}}\psi_{_{h_d}}.
\label{eq14}
\end{eqnarray}
One can then deduce that Eq. (\ref{eq13}) is a mass term of the Dirac field 
$\tilde{h}^T=\left(\psi_{_{H_u}},
\overline{\psi}_{_{h_d}}\right)$. This heavy higgsino field, with a mass $\mu_H$,
is integrated out in the effective theory.

\section{The effective Lagrangian with dimension-odd operators
\label{sec3}}
\indent\indent

In the effective theory that results when the heavy fields are integrated out 
neutrino masses are in leading order generated by the following  dimension-five
operators consisting of the SM fields \cite{Weinberg}:
\begin{eqnarray}
&&\delta{\cal L}_{d=5}={1\over2}(C_{_1})_{_{J,I}}\Big(\overline{(e_{_L}^J)^c}
\;\tilde{\Phi}^*\Big)\Big(\tilde{\Phi}^\dagger\; e_{_L}^I\Big)
\nonumber\\&&\hspace{0.2cm}
+{1\over2}(C_{_2})_{_{J,I}}\Big(\overline{(e_{_L}^J)^c}\sigma^a\tilde{\Phi}^*\Big)
\Big(\tilde{\Phi}^\dagger\sigma^a e_{_L}^I\Big)+h.c.,
\label{eq17}
\end{eqnarray}
where $(C_{_1})_{_{J,I}},(C_{_2})_{_{J,I}}$ are Wilson coefficients.
\begin{figure}[t]
\setlength{\unitlength}{1mm}
\begin{center}
\begin{picture}(0,20)(-10,-5)
\put(-62,-100){\includegraphics{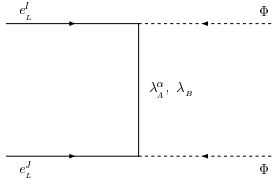}}
\end{picture}
\caption[]{The tree level diagrams that lead to the dimension-five
operators.} 
\label{fig1}
\end{center}
\end{figure}
The tree-level Feynman diagrams inducing the above effective
Lagrangian are given in Fig. \ref{fig1}. The Wilson coefficients corresponding to 
these diagrams are
\begin{eqnarray}
&&(C_{_1})_{_{J,I}}={g_1^2c_{_\beta}^2\over|m_{_1}|}\Lambda_{_{0,I}}
\Lambda_{_{0,J}}e^{-i\theta_1}\;,\nonumber\\
&&(C_{_2})_{_{J,I}}={g_2^2c_{_\beta}^2\over|m_{_2}|}\Lambda_{_{0,I}}
\Lambda_{_{0,J}}e^{-i\theta_2}\;,
\label{eq18}
\end{eqnarray}
where $\tan\beta=\upsilon_{_u}/\upsilon_{_d},\;c_{_\beta}=\cos\beta,\;
s_{_\beta}=\sin\beta$, and the parameters $\Lambda_{I,J}$ are definited as
\begin{eqnarray}
&&\Lambda_{_{0,0}}={\upsilon_{_0}|\epsilon_{_0}|\over\upsilon_{_d}\mu_{_H}}
+\sum\limits_{I=1}^3{\upsilon_{_I}|\epsilon_{_I}|
\over\upsilon_{_d}\mu_{_H}}e^{-i\varphi_{_I}}\;,\nonumber\\
&&\Lambda_{_{I,0}}={\upsilon_{_{d_{(I-1)}}}|\epsilon_{_I}|
\over\upsilon_{_{d_{I}}}\mu_{_H}}e^{-i\varphi_{_I}}
-\sum\limits_{J=0}^{I-1}{\upsilon_{_I}\upsilon_{_J}
|\epsilon_{_J}|\over\upsilon_{_{d_{(I-1)}}}\upsilon_{_{d_{I}}}
\mu_{_H}}e^{-i\varphi_{_J}}\;,\nonumber\\
&&\Lambda_{_{0,I}}={\upsilon_{_I}\mu_{_{H_{(I-1)}}}
\over\upsilon_{_d}\mu_{_{H_I}}}-\sum\limits_{J=0}^{I-1}
{|\epsilon_{_I}\epsilon_{_J}|e^{i(\varphi_{_I}-\varphi_{_\beta})}
\upsilon_{_J}\over\upsilon_{_d}\mu_{_{H_{(I-1)}}}\mu_{_{H_I}}}
\label{eq19}
\end{eqnarray}
with $I,J=1,\;2,\;3$.
When the EW symmetry breaks down, one ends up with the following mass matrix for 
neutrinos:
\begin{eqnarray}
&&{\bf m}_{_\nu}={1\over2}c_{_\beta}^2\upsilon^2\Big({g_1^2e^{-i\theta_1}\over|m
_{_1}|}
+{g_2^2e^{-i\theta_2}\over|m_{_2}|}\Big)
\nonumber\\
&&\hspace{1.0cm}\times
\left(\begin{array}{ccc}\Lambda_{_{0,1}}^2&
\Lambda_{_{0,1}}\Lambda_{_{0,2}}&\Lambda_{_{0,1}}\Lambda_{_{0,3}}\\ \\
\Lambda_{_{0,1}}\Lambda_{_{0,2}}&\Lambda_{_{0,2}}^2&\Lambda_{_{0,2}}\Lambda_{_{0,3}}\\ \\
\Lambda_{_{0,1}}\Lambda_{_{0,3}}&\Lambda_{_{0,2}}\Lambda_{_{0,3}}
&\Lambda_{_{0,3}}^2\end{array}\right)\;.
\label{eq20}
\end{eqnarray}
This matrix is diagonalized by the unitary matrix 
\begin{eqnarray}
&&V=\left(\begin{array}{ccc}0&-{A_{_1}\over A}&{\Lambda_{_{0,1}}^*\over A}\\ \\
-{\Lambda_{_{0,3}}\over A}&{\Lambda_{_{0,1}}\Lambda_{_{0,2}}^*\over AA_{_1}}&
{\Lambda_{_{0,2}}^*\over A}\\ \\
{\Lambda_{_{0,2}}\over A}&{\Lambda_{_{0,1}}\Lambda_{_{0,3}}^*\over AA_{_1}}
&{\Lambda_{_{0,3}}^*\over A}\;,
\end{array}\right)
\label{eq21}
\end{eqnarray}
where 
\begin{eqnarray}
&&A=\sqrt{|\Lambda_{_{0,1}}|^2+|\Lambda_{_{0,2}}|^2+|\Lambda_{_{0,3}}|^2}\;,
\nonumber\\
&&A_{_1}=\sqrt{|\Lambda_{_{0,1}}|^2+|\Lambda_{_{0,2}}|^2}\;,
\label{eq21-a}
\end{eqnarray}
giving
\begin{eqnarray}
&&V^\dagger{M}_{_\nu}V={1\over4}c_{_\beta}^2\upsilon^2
\Big({g_1^2e^{-i\theta_1}\over|m_{_1}|}+{g_2^2e^{-i\theta_2}\over|m_{_2}|}\Big)
\nonumber\\
&&\hspace{1.0cm}\times
{\rm diag}(0,\;0,\;A^2)\;.
\label{eq22}
\end{eqnarray}
Hence only one neutrino flavor acquires a non-vanishing mass at tree level.
This mass is given more explicitly by
\begin{eqnarray}
&&m_{_{\nu_3}}={1\over2\mu_{_H}^2}\Big|{g_1^2e^{-i\theta_1}\over|m_{_1}|}
+{g_2^2e^{-i\theta_2}\over|m_{_2}|}\Big|
\nonumber\\
&&\hspace{1.0cm}\times
\Big[\upsilon_{_d}^2\mu_{_H}^2
-\Big|\upsilon_{_0}|\epsilon_{_0}|+\sum\limits_{I=1}^3\upsilon_{_I}
|\epsilon_{_I}|e^{i\varphi_{_I}}\Big|^2\Big]\;.
\label{eq23}
\end{eqnarray}
By setting the phases zero and taking account of the approximations used, we find 
our result coincident with that of Ref. \cite{Hempfling}.

\begin{figure}[t]
\setlength{\unitlength}{1mm}
\begin{center}
\begin{picture}(0,30)(-10,-15)
\put(-62,-100){\includegraphics{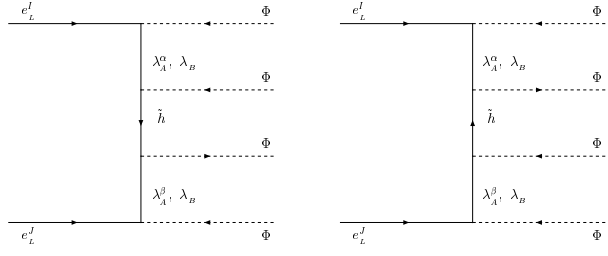}}
\end{picture}
\caption[]{The tree level diagrams that lead to the dimension-seven
operators.} 
\label{fig2}
\end{center}
\end{figure}

An obvious conclusion  from above is that the neutrino oscillation data, which 
indicates the existence of two distinctly different mass-difference scales, is not 
explained in the MSSM at tree level. Therefore, a study of higher order 
contributions to the neutrino mass Lagrangian is necessary. 

Let us first consider the  corrections of dimension-seven operators to the neutrino 
mass matrix. The tree-level Wilson coefficients of the dimension-seven operatos and 
the Wilson coefficients of the dimension-five operators (\ref{eq17}) at one-loop 
level belong to the same order of perturbative expansion.  The dimension-seven 
operators that give corrections to the neutrino
mass matrix are the following:
\begin{eqnarray}
&&\delta{\cal L}_{d=7}={1\over2}(C_{_3})_{_{J,I}}\Big(\Phi^\dagger\Phi\Big)
\Big(\overline{(e_{_L}^J)^c}\;\tilde{\Phi}^*\Big)
\Big(\tilde{\Phi}^\dagger\; e_{_L}^I\Big)
\nonumber\\&&\hspace{0.2cm}
+{1\over2}(C_{_4})_{_{J,I}}\Big(\Phi^\dagger\sigma^a\Phi\Big)
\Big(\overline{(e_{_L}^J)^c}\tilde{\Phi}^*\Big)
\Big(\tilde{\Phi}^\dagger\sigma^a e_{_L}^I\Big)
\nonumber\\&&\hspace{0.2cm}
+{1\over2}(C_{_5})_{_{J,I}}\Big(\Phi^\dagger\sigma^a\Phi\Big)
\Big(\overline{(e_{_L}^J)^c}\sigma^a \tilde{\Phi}^*\Big)
\Big(\tilde{\Phi}^\dagger e_{_L}^I\Big)
\nonumber\\&&\hspace{0.2cm}
+{1\over2}(C_{_6})_{_{J,I}}\Big(\Phi^\dagger\sigma^a\sigma^b\Phi\Big)
\Big(\overline{(e_{_L}^J)^c}\sigma^a \tilde{\Phi}^*\Big)
\Big(\tilde{\Phi}^\dagger\sigma^b e_{_L}^I\Big)
\nonumber\\&&\hspace{0.2cm}
+h.c.\;,
\label{eq24}
\end{eqnarray}
where the Wilson coefficients have, refering to  Fig. \ref{fig2}, the expressions
\begin{eqnarray}
&&(C_{_3})_{_{J,I}}={g_1^4\over|m_{_1}|^2\mu_{_H}}s_{_\beta}c_{_\beta}^3
e^{-i(\theta_{_\mu}+2\theta_{_1})}
\nonumber\\
&&\hspace{1.5cm}\times
\Lambda_{_{0,0}}\Lambda_{_{0,I}}\Lambda_{_{0,J}}
\;,\nonumber\\
&&(C_{_4})_{_{J,I}}={g_1^2g_2^2\over|m_{_1}m_{_2}|\mu_{_H}}s_{_\beta}c_{_
\beta}^3
e^{-i(\theta_{_\mu}+\theta_{_1}+\theta_{_2})}
\nonumber\\
&&\hspace{1.5cm}\times
\Lambda_{_{0,0}}\Lambda_{_{0,I}}\Lambda_{_{0,J}}
\;,\nonumber\\
&&(C_{_5})_{_{J,I}}=(C_{_4})_{_{J,I}}
\;,\nonumber\\
&&(C_{_6})_{_{J,I}}={g_2^4\over|m_{_2}|^2\mu_{_H}}s_{_\beta}c_{_\beta}^3
e^{-i(\theta_{_\mu}+2\theta_{_2})}
\nonumber\\
&&\hspace{1.5cm}\times
\Lambda_{_{0,0}}\Lambda_{_{0,I}}\Lambda_{_{0,J}}\;.
\label{eq25}
\end{eqnarray}
After the electroweak symmetry breaking these operators yield corrections to the
neutrino mass matrix, which are accounted for by the following replacement in Eq. 
(\ref{eq20}):
\begin{eqnarray}
&&\Big({g_1^2e^{-i\theta_1}\over|m_{_1}|}
+{g_2^2e^{-i\theta_2}\over|m_{_2}|}\Big)
\nonumber\\
&&\longrightarrow\Big({g_1^2e^{-i\theta_1}\over|m_{_1}|}
+{g_2^2e^{-i\theta_2}\over|m_{_2}|}\Big)\Big[1
-{1\over2}s_{_\beta}c_{_\beta}\Lambda_{_{0,0}}
\nonumber\\
&&\hspace{0.5cm}\times
{\upsilon^2\over\mu_{_H}}e^{-i\theta_{_\mu}}
\Big({g_1^2e^{-i\theta_1}\over|m_{_1}|}+{g_2^2e^{-i\theta_2}\over|m_{_2}|}\Big)
\Big]\;.
\label{eq26}
\end{eqnarray}
The same replacement should be done in Eq. (\ref{eq23}). 
We realize that when the tree-level contributions of the dimension-seven operators 
taken account, still only one neutrino flavour acquires a nonzero
mass.

\begin{figure}[t]
\setlength{\unitlength}{1mm}
\begin{center}
\begin{picture}(0,130)(-10,-90)
\put(-62,-100){\includegraphics{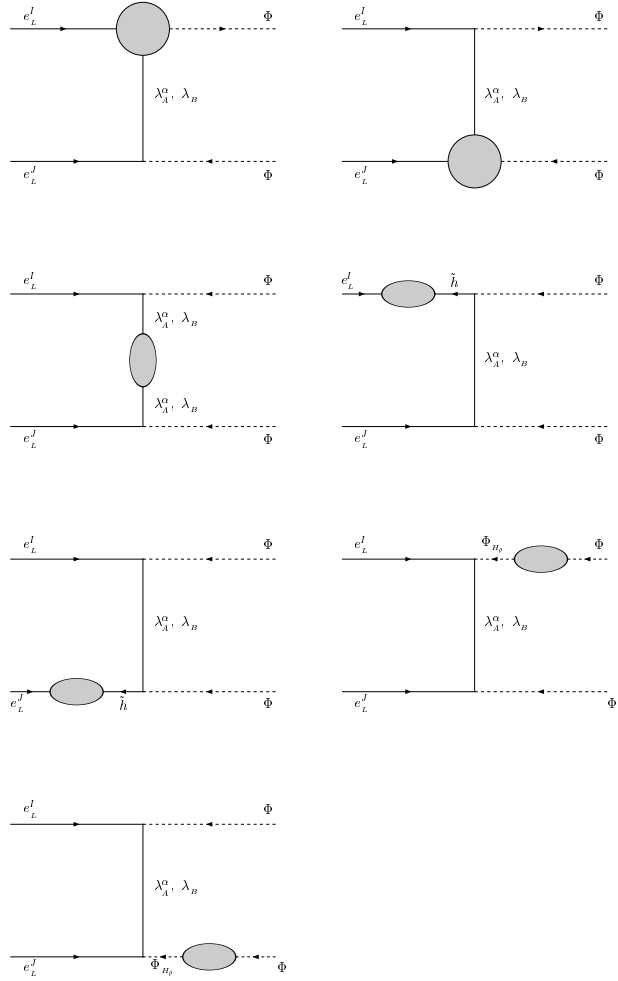}}
\end{picture}
\caption[]{The one-loop diagrams that lead to the dimension-five
operators after the matching. The triangle, and self-energy diagrams denoted by 
blobs are
presented in Figs. 5 to 8.} 
\label{fig3}
\end{center}
\end{figure}

\begin{figure}[t]
\setlength{\unitlength}{1mm}
\begin{center}
\begin{picture}(0,30)(-10,10)
\put(-62,-100){\includegraphics{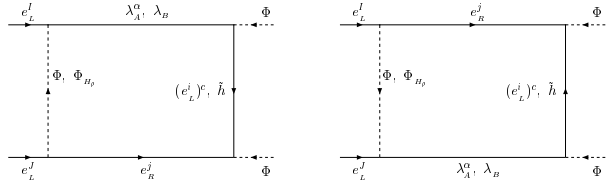}}
\end{picture}
\caption[]{The box-diagrams that lead to the dimension-five
operators after the matching.} 
\label{fig4}
\end{center}
\end{figure}

In Fig. \ref{fig3} and Fig. \ref{fig4} we present the one-loop diagrams that 
contribute to the Wilson coefficients of dimension-five operator
after the matching procedure. In the full theory, the self-energy
corrections to external leg, appearing in the diagrams in Fig. \ref{fig3}, should 
in princible include also the light-light contributions, such as
$e_{_L}e_{_L},\;\Phi\Phi$. However, such diagrams would not give any contribution
to the dimension-five operators after the matching of the full and 
effective theory \cite{feng2}. A similar conclusion is true also for the
vertex correction diagrams that involve only the light degrees of freedom.

The lengthy expressions of the ensuing Wilson coefficients are presented 
in appendix \ref{app1}. 

\begin{figure}[t]
\setlength{\unitlength}{1mm}
\begin{center}
\begin{picture}(0,130)(-10,-90)
\put(-62,-100){\includegraphics{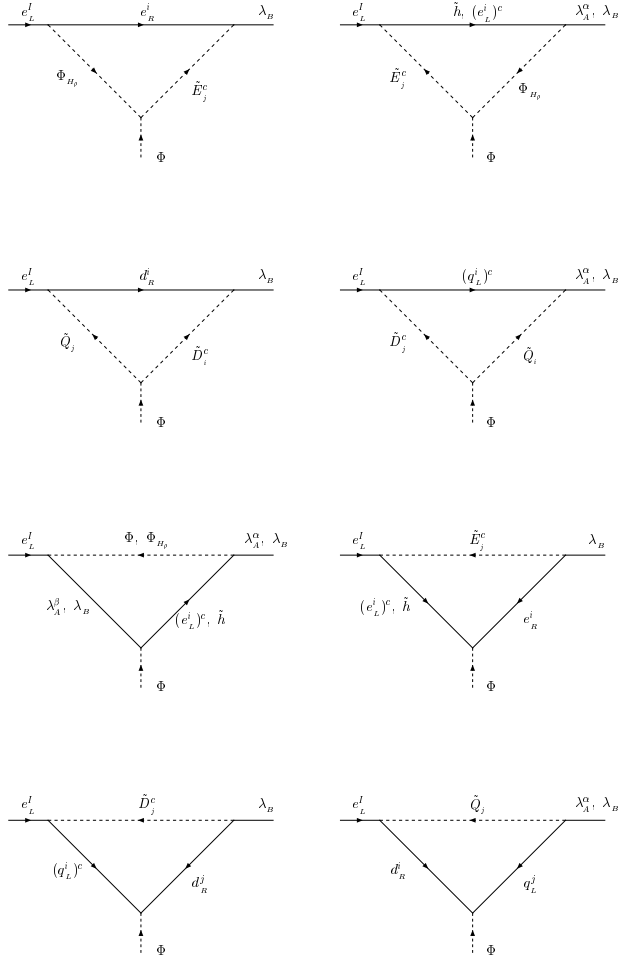}}
\end{picture}
\caption[]{The one-loop diagrams that lead to the dimension-five
operators after the matching. The triangle-, and self energy-diagrams are
presented in Figs. 5 to 7.} 
\label{fig5}
\end{center}
\end{figure}
\begin{figure}
\setlength{\unitlength}{1mm}
\begin{center}
\begin{picture}(0,60)(-10,-15)
\put(-62,-100){\includegraphics{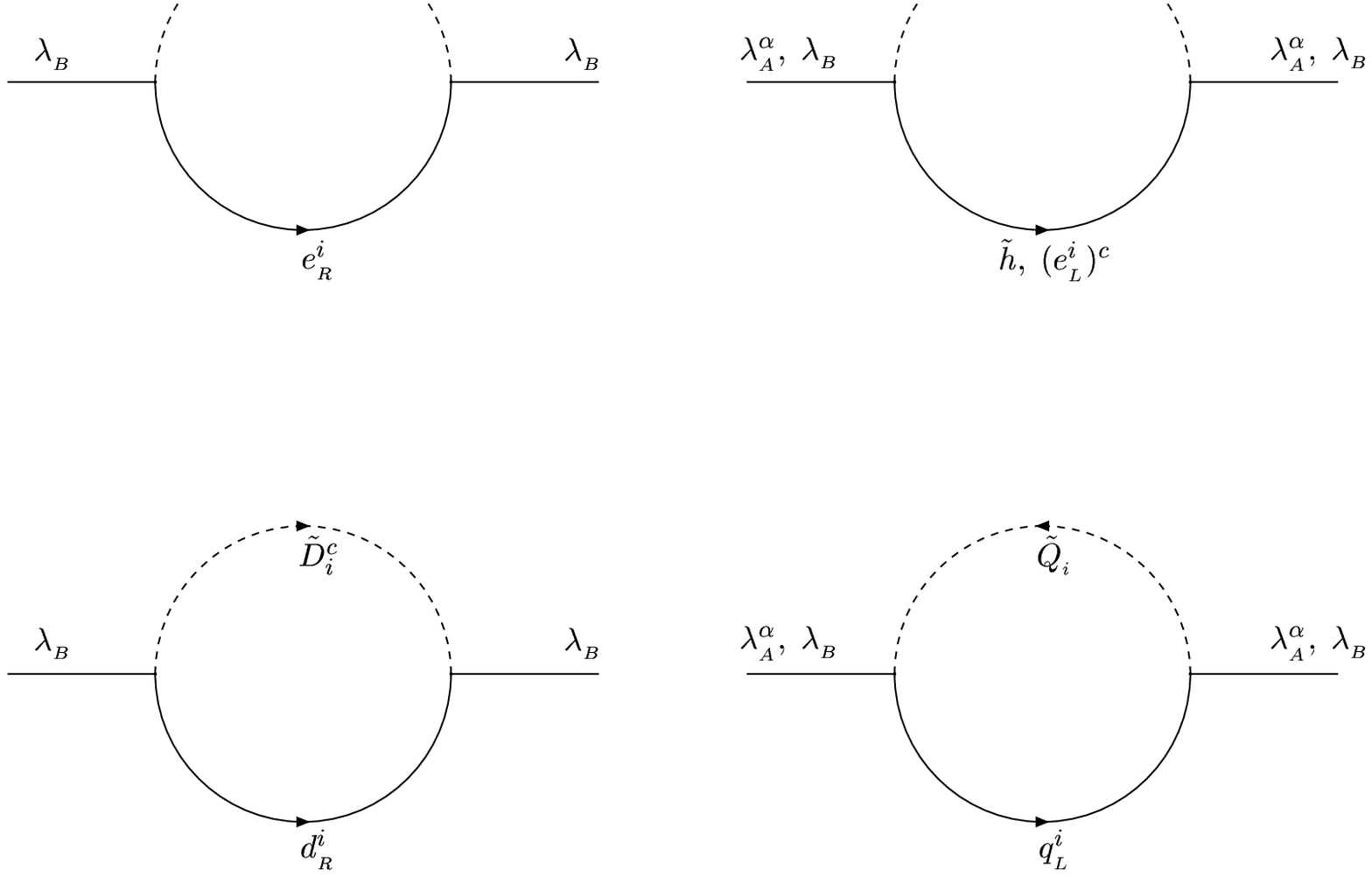}}
\end{picture}
\caption[]{The gaugino self-energy diagrams that lead to the dimension-five
operators after the matching.} 
\label{fig6}
\end{center}
\end{figure}
\begin{figure}
\setlength{\unitlength}{1mm}
\begin{center}
\begin{picture}(0,90)(-10,-50)
\put(-62,-100){\includegraphics{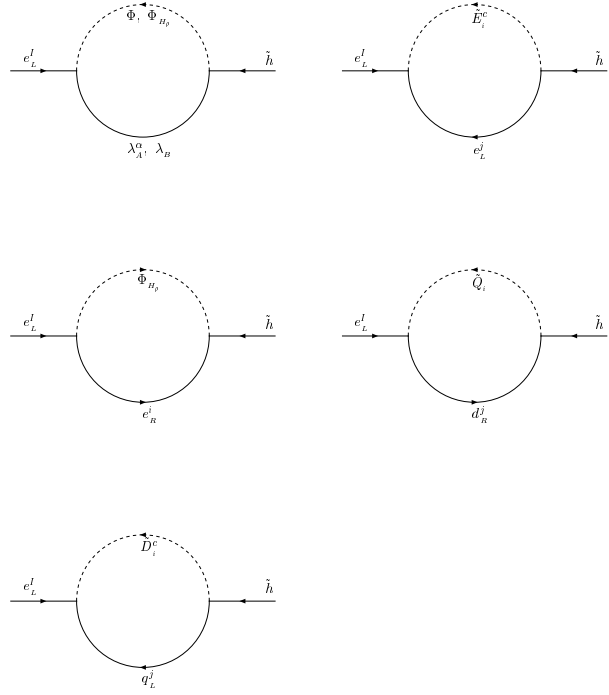}}
\end{picture}
\caption[]{The lepton higgsino self-energy diagrams that lead to the dimension-five
operators after the matching.} 
\label{fig7}
\end{center}
\end{figure}
\begin{figure}
\setlength{\unitlength}{1mm}
\begin{center}
\begin{picture}(0,90)(-10,-50)
\put(-62,-100){\includegraphics{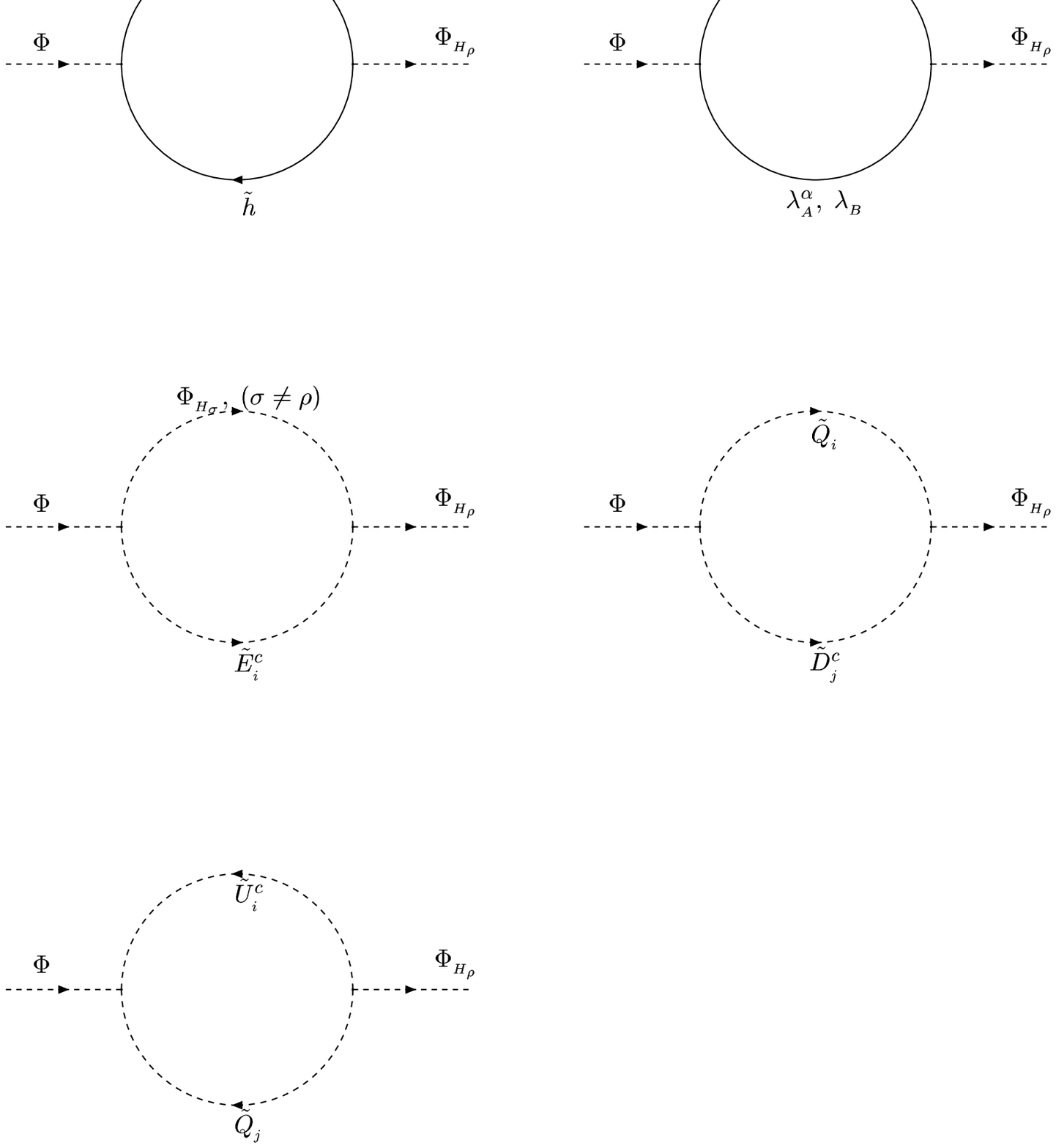}}
\end{picture}
\caption[]{The Higgs self-energy diagrams that lead to the dimension-five
operators after the matching.} 
\label{fig8}
\end{center}
\end{figure}

After the EW symmetry breaking, the neutrino
mass matrix with the one-loop corrections included can be written as
\begin{eqnarray}
&&{M}_{_\nu}+\delta{M}_{_\nu},
\label{eq27}
\end{eqnarray}
where ${M}_{_\nu}$ denotes the contribution from the tree level
dimension-five and dimension-seven operators and $\delta{M}_{_\nu}$ is the 
correction
from the dimension-five operators at the one-loop level, given by
\begin{eqnarray}
&&\Big(\delta{M}_{_\nu}\Big)_{_{J,I}}={\upsilon^2\over2}\Big(\delta C_{_1}^{(1)}
+\delta C_{_2}^{(1)}+\delta C_{_1}^{(2)}
\nonumber\\
&&\hspace{2.cm}
+\delta C_{_2}^{(2)}\Big)_{_{J,I}}\;,
\label{eq28}
\end{eqnarray}
where the quantities $\delta C_{_{1,2}}^{(1,2)}$ are defined in 
appendix \ref{app1}.

The neutrino mass matrix is diagonalized, as usual,  by a unitary transformation
\begin{eqnarray}
U^\dagger\Big({M}_{_\nu}+\delta{M}_{_\nu}\Big)U
={\rm diag}(m_{_{\nu_1}},\;m_{_{\nu_2}},\;m_{_{\nu_3}})\;,
\label{eq29}
\end{eqnarray}
where the $3\times3$ unitary matrix $U$ generally contains 3 rotation angles and 
three physical phases ($CP$-phases), and $m_{_{\nu_i}}$ denote the real positive 
mass eigenvalues of the
light Majorana neutrinos. Among the three $CP$ phases, two are so-called Majorana 
phases associated with lepton number violating  terms in the mass Lagrangian and 
the third one is a so-called Dirac phase. One can factorize the Majorana phases by 
defining   the transformation matrix $U$ as
\begin{eqnarray}
&&U=U_{_{L}}D_{_m}\;,
\label{eq30}
\end{eqnarray}
where $U_{_{L}}$ is the lepton mixing matrix, a counter part of the 
Cabibbo-Kobayashi-Maskawa mixing matrix of quarks, containing 
the Dirac phase, and $D_{_m}$ is a diagonal matrix with two
Majorana phases. One can use the following parametrization:
\begin{eqnarray}
&&D_{_m}=\left(\begin{array}{ccc}e^{i\omega_{_1}}&0&0\\
0&e^{i\omega_{_2}}&0\\0&0&1\end{array}\right)\;,
\label{eq31}
\end{eqnarray}
\begin{widetext}
\begin{eqnarray}
&&U_{_{CKM}}=\left(\begin{array}{ccc}c_{_{12}}c_{_{13}}&s_{_{12}}c_{_{13}}&
s_{_{13}}e^{-i\delta}\\
-s_{_{12}}c_{_{23}}-c_{_{12}}s_{_{23}}s_{_{13}}e^{i\delta}&
c_{_{12}}c_{_{23}}-s_{_{12}}s_{_{23}}s_{_{13}}e^{i\delta}&s_{_{23}}c_{_{1
3}}\\
s_{_{12}}s_{_{23}}-c_{_{12}}c_{_{23}}s_{_{13}}e^{i\delta}&
-c_{_{12}}s_{_{23}}-s_{_{12}}c_{_{23}}s_{_{13}}e^{i\delta}&c_{_{23}}c_{_{13}}
\end{array}\right)\;,
\label{eq32}
\end{eqnarray}
\end{widetext}
with $c_{_{ij}}=\cos\theta_{_{ij}},\;s_{_{ij}}=\sin\theta_{_{ij}}$ and
$\theta_{_{ij}}$ denote the mixing angles. 

The  rotation angles $\theta_{ij}$ and the phases $\delta$ and $\omega_i$ appearing 
in the lepton mixing matrix $U_L$, as well as the values of neutrino masses, can be 
expressed in terms of the parameters of the the model. In the following we will 
study the constraints set by the atmospheric and solar neutrino oscillation data on 
the model parameters.

\section{Constraints from oscillation data}
\indent\indent
The neutrino oscillation data have outlined
the mass and mixing pattern of neutrinos. According to a global fit to the data, 
the 
oscillations are best explained in terms of the following set of parameters
\cite{Gonzalez-Garcia}
\begin{eqnarray}
&&\Delta m_{_{32}}^2=2.0\times10^{-3}\;{\rm eV}^2,\nonumber\\
&&\Delta m_{_{21}}^2=7.2\times10^{-5}\;{\rm eV}^2,\nonumber\\
&&\sin^2\theta_{_{23}}=0.5,\;\sin^2\theta_{_{12}}=0.3,\nonumber\\
&&\sin^2\theta_{_{13}}<0.06\;.
\label{eq34}
\end{eqnarray}

Let us now discuss the implications of these empirical results for the parameters 
of the $/\!\!\!\!R$MSSM.
Beside the parameters appearing in the MSSM, the $/\!\!\!\!R$MSSM includes
generally many new ones (in general complex) related to lepton number violation:
three bilinear ($\epsilon_{_I}$) and thirty six trilinear (nine $ 
\lambda_{_{IJK}}$, twenty seven $\lambda_{_{IJK}}^\prime$) couplings 
in the superpotential, together with
the corresponding soft breaking (three $B_{_I}$) and A-terms 
(nine $A_{_{IJK}}$, twenty seven 
$A_{_{IJK}}^\prime$) and three additional $/\!\!\!\!R$ soft masses 
$(m_{_{\tilde L}}^2)_{_{0I}}$. There are only six physical parameters
among the nine bilinear $/\!\!\!\!R$ parameters ($\epsilon_{_I},\;B_{_I}
,\;(m_{_{\tilde L}}^2)_{_{0I}}$) due to the ambiguity in the choice
of the ($H_{_d},\;{\tilde L}$) basis. It is obvious, giving the large number of 
parameters, the oscillation data alone would leave in the general case much free 
space for the choice of parameters. In the following we will not consider the most 
general situation but restrict ourselves to the case where only the bilinear 
$/\!\!\!\!R$ couplings are present. We are particularly interested in the effects 
of the complex phases associated with these couplings.

Neutrino masses in the framework of  $/\!\!\!\!R$MSSM are extensively discussed in 
the literature, but the studies are usually restricted to the $CP$-conserving case 
leaving the possible effects of $CP$ phases with less attention. 
In the MSSM, the soft broken terms provide new sources for the $CP$
violation, in addition to the CKM mechanism of the Standard Model.
At present, the strictest constraints on the $CP$ 
phases originate from the experimental bounds on the electric dipole moments (EDM) 
of the electron and the neutron. Nevertheless,
if one invokes a cancellation mechanism among different supersymmetric
contributions \cite{edm1} or choose the sfermions of two generations
heavy enough \cite{edm2}, the loop-induced EDM's yield  a
bound only for the argument of the parameter $\mu$, implying  $\mu\le 
\pi/(5\tan\beta)$,
leaving the other explicit $CP$-violation phases unconstrained.

In principle, all the parameters associated with $R$-parity violation are complex
in the model we are considering. For simplicity, we will now assume that all 
parameters are real except the gaugino masses. 
Furthermore, in order to apply the effective theory method safely,
we assume the new physics scale is $\mu_{_{NP}}=10\;{\rm TeV}$, in other words, 
much higher than the electroweak scale. We choose
the basis $\upsilon_{_I}=0\;(I=1,\;2,\;3)$, and we set 
\begin{eqnarray}
&&(m_{_{\tilde L}})_{_{IJ}}
={\rm diag}(1,\;\sqrt{2},\;\sqrt{3})\times10\;{\rm TeV},\nonumber\\
&&(m_{_{\tilde L}})_{_{0I}}=(100,\;100,\;100)\;{\rm GeV},\nonumber\\
&&(m_{_{\tilde E}})_{_{IJ}}=(m_{_{\tilde U}})_{_{IJ}}
=(m_{_{\tilde D}})_{_{IJ}}={\rm diag}(5,\;5,\;1)\;{\rm TeV},\nonumber\\
&&(m_{_{\tilde Q}})_{_{IJ}}={\rm diag}(5,\;5,\;1)\;{\rm TeV},\nonumber\\
&&A_{_{IJ}}^u=A_{_{IJ}}^d=A_{_{IJ}}^e=0\;,
(I,\;J=1,\;2,\;3)
\;.\nonumber
\end{eqnarray}
For the bilinear couplings we use the values
\begin{eqnarray}
&&\epsilon_{_0}=5\;{\rm TeV},\;\epsilon_{_1}=0.01\;{\rm GeV},\nonumber\\
&&\epsilon_{_2}=0.3\;{\rm GeV},\;\epsilon_{_3}=4\;{\rm GeV},\nonumber\\
&&B_{_0}=100\;{\rm TeV}^2,\;B_{_1}=B_{_2}=B_{_3}=100\;{\rm GeV}^2\;.\nonumber\\
\end{eqnarray}
In order to obtain small neutrino masses, $m_{_\nu}< 0.1\;{\rm eV}$,
we assume the gaugino masses very large, $|m_{_1}|=|m_{_2}|=100\;{\rm TeV}$.
We consider the numerical values of the parameters given above to be conceivable 
and representative for the model.
\begin{figure}[t]
\setlength{\unitlength}{1mm}
\begin{center}
\begin{picture}(0,120)(-10,-90)
\put(-62,-100){\includegraphics{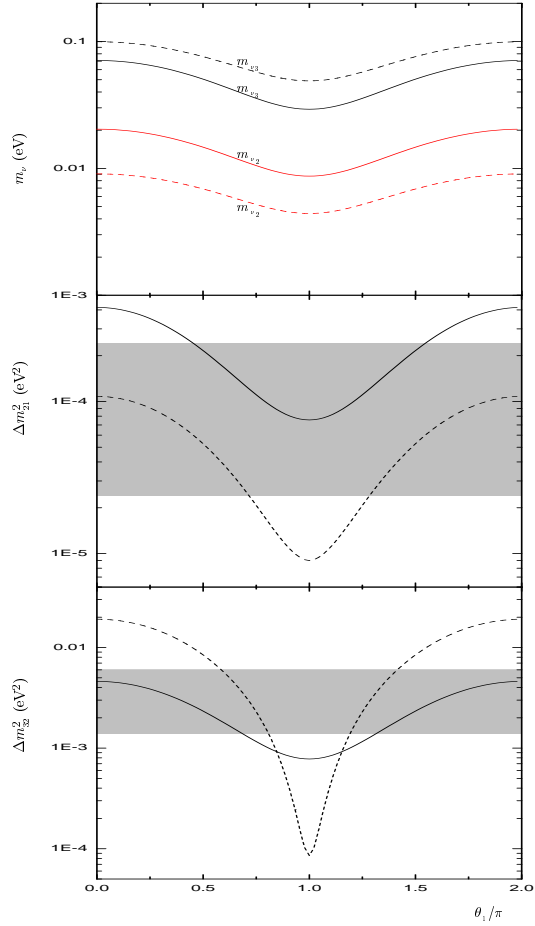}}
\end{picture}
\caption[]{The neutrino mass-squared
differences $\Delta m_{_{32}}^2,\;\Delta m_{_{21}}^2$, and the neutrino 
masses $m_{_{\nu_2}},\;m_{_{\nu_3}}$ versus the $CP$ phase 
$\theta_{_1}=arg(m_{_1})$ for $\theta_{_2}=arg(m_{_2})=0$.
The solid lines correspond $\tan\beta=50$ and the dash lines to $\tan\beta=5$.
The gray bands represent the region allowed by the global fit of the solar 
and atmospheric neutrino data  at $3\sigma$ level. For the values of the
other parameters, see the text.} 
\label{fig9}
\end{center}
\end{figure}
\begin{figure}[t]
\setlength{\unitlength}{1mm}
\begin{center}
\begin{picture}(0,120)(-10,-90)
\put(-62,-100){\includegraphics{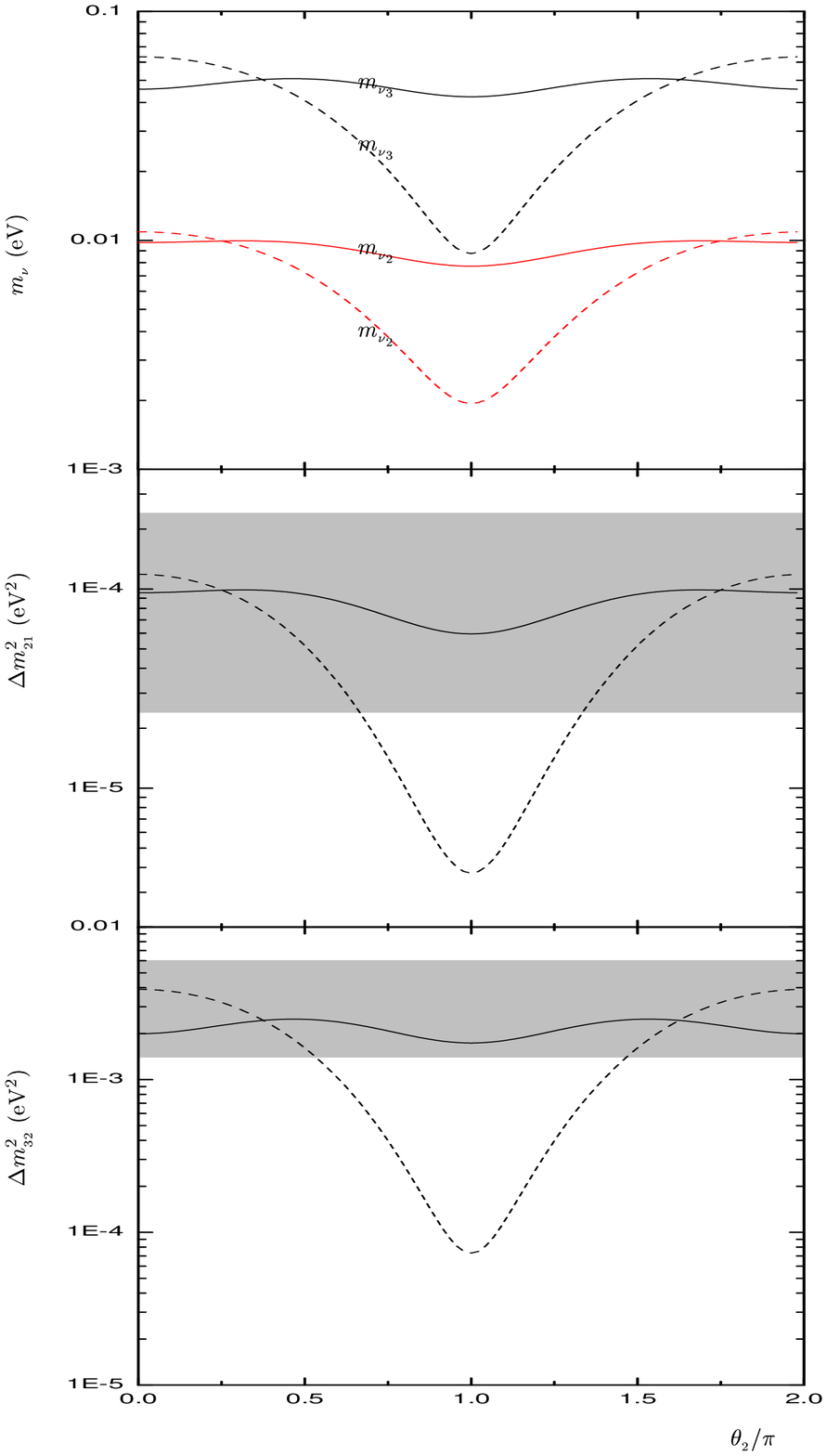}}
\end{picture}
\caption[]{The neutrino mass-squared
differences $\Delta m_{_{32}}^2,\;\Delta m_{_{21}}^2$, and the neutrino 
masses $m_{_{\nu_2}},\;m_{_{\nu_3}}$ versus the $CP$ phase 
$\theta_{_2}=arg(m_{_2})$ for $\theta_{_1}=arg(m_{_1})=0$.
The solid lines correspond to $\tan\beta=50$, the dash lines $\tan\beta=5$.} 
\label{fig10}
\end{center}
\end{figure}

In order to demonstrate the effects of $CP$-phases on the masses of neutrinos we 
plot in Fig.\ref{fig9}, in the case $\theta_{_2}=arg(m_{_2})=0$,   the neutrino 
mass-squared
differences $\Delta m_{_{32}}^2,\;\Delta m_{_{21}}^2$, and the neutrino 
masses $m_{_{\nu_2}},\;m_{_{\nu_3}}$ versus the $CP$-phase 
$\theta_{_1}=arg(m_{_1})$. We have considered the situation for both large 
($\tan\beta=50$, solid line) and small ($\tan\beta=5$, dashed line) values of the 
ratio $\tan\beta=\upsilon_u/\upsilon_d$. In Fig.\ref{fig10} we present a similar 
plot the same for $\theta_{_1}=arg(m_{_1})=0$ and a varying 
$\theta_{_2}=arg(m_{_2})$. It is seen from these plots that the solar 
and atmospheric neutrino data, which allow the $3\sigma$-range shown in gray in the 
figures, sets non-trivial connstraints on the complex phases, in particular in the 
case of a small $\tan\beta$.

Let us finally note that we have found, by scanning the parameter space, that when 
all the trilinear $/\!\!\!\!L$ couplings together with corresponding
soft terms are included, it is possible to realize in this model a situation, where 
a large neutrino mass hierarchy is associated with a large mixing. In general, this 
kind of situation is not so easy to achieve in a definite theoretical framework.

\section{Summary\label{sec4}}
\indent\indent
There are many ways to generate neutrino masses in the Minimal Supersymmetric 
Standard Model without R-parity. In this paper we have derived, using the effective 
field theory approach, the high-dimension operators relevant for neutrino masses, 
which the R-parity breaking couplings induce. Among these
high-dimension operators, the dimension-odd operators originating from
the lepton number violation couplings will lead to nonzero 
Majorana masses after the EW symmetry breaking. By properly defining the 
heavy fields, we have derived the Wilson coefficients of dimension-five operators 
at one loop level and the those of dimension-seven operators at tree level. We have 
demonstrated the effect of $CP$ phases on neutrino masses by looking at a specific 
case, where the only complex $R$-parity violating couplings are the soft bilinear 
gaugino mass terms.

\begin{acknowledgments}

The work has been supported by the Academy of Finland under the contracts 
no.\ 104915 and 107293.
\end{acknowledgments}

\appendix
\section{The one-loop corrections on the Wilson coefficients of the dimension-five 
operators\label{app1}}
\indent\indent
In this appendix, we present the one-loop corrections on the Wilson coefficients 
of the dimension-five operators that are present in Eq. (\ref{eq17}). The triangle 
diagram
corrections are
\begin{widetext}
\begin{eqnarray}
&&(\delta C_{_1}^{(1)})_{_{J,I}}={g_1^2\mu_{_H}\over2\mu_{_{\rm NP}}^2|m_{_1}|}
e^{-i(\theta_{_\mu}+\theta_{_1})}
c_{_\beta}^2\Big[\Lambda_{_{0,J}}({\bf Y}_{_{R_M}})_{_{0,I}}
+\Lambda_{_{0,I}}({\bf Y}_{_{R_M}})_{_{0,J}}\Big]
({\cal Z}_{_R})_{_{M,i}}({\cal Z}_{_R}^\dagger)_{_{i,N}}
({\bf A}_{_{S_N}})_{_{0,\alpha}}
\nonumber\\
&&\hspace{2.2cm}\times
({\cal Z}_{_H})_{_{\alpha,1}}
C(x_{_{\tilde h}},x_{_{H_{\alpha}}},x_{_{R_i}})
\nonumber\\
&&\hspace{2.2cm}
+{g_1^4\over2|m_{_1}|}e^{-i\theta_{_1}}c_{_\beta}^2\Lambda_{_{0,I}}
\Lambda_{_{0,J}}\Big[s_{_\beta}^2B(x_{_{\tilde h}},x_{_{\tilde B}})
-s_{_\beta}c_{_\beta}\Lambda_{_{0,0}}e^{-i(\theta_{_\mu}+\theta_{_1})}
\sqrt{x_{_{\tilde h}}x_{_{\tilde B}}}P(x_{_{\tilde h}},x_{_{\tilde B}})\Big]
\nonumber\\
&&\hspace{2.2cm}
+{3g_1^2g_2^2\over2|m_{_1}|}e^{-i\theta_{_1}}c_{_\beta}^2\Lambda_{_{0,I}}
\Lambda_{_{0,J}}\Big[s_{_\beta}^2B(x_{_{\tilde h}},x_{_{\tilde A}})
-s_{_\beta}c_{_\beta}\Lambda_{_{0,0}}e^{-i(\theta_{_\mu}+\theta_{_2})}
\sqrt{x_{_{\tilde h}}x_{_{\tilde A}}}P(x_{_{\tilde h}},x_{_{\tilde A}})\Big]
\nonumber\\
&&\hspace{2.2cm}
+{g_1^4\over4|m_{_1}|}e^{-i\theta_{_1}}c_{_\beta}^2\Big[
2s_{_\beta}\Lambda_{_{0,I}}\Lambda_{_{0,J}}({\cal Z}_{_H}^\dagger)_{_{1,\alpha}}
+\sum\limits_{i=1}^3\Big(\Lambda_{_{0,I}}\Lambda_{_{i,J}}
+\Lambda_{_{i,I}}\Lambda_{_{0,J}}\Big)
({\cal Z}_{_H}^\dagger)_{_{i+1,\alpha}}\Big]
\nonumber\\
&&\hspace{2.2cm}\times
({\cal Z}_{_H})_{_{\alpha,1}}
\Big[s_{_\beta}C_0(x_{_{H_{\alpha}}},x_{_{\tilde h}},x_{_{\tilde B}})
\nonumber\\
&&\hspace{2.2cm}
-c_{_\beta}\Lambda_{_{0,0}}\sqrt{x_{_{\tilde h}}x_{_{\tilde B}}}
e^{-i(\theta_{_\mu}+\theta_{_1})}C(x_{_{H_{\alpha}}},x_{_{\tilde h}},x_{_{\tilde 
B}})\Big]
+{3g_1^2g_2^2\over4|m_{_1}|}e^{-i\theta_{_1}}c_{_\beta}^2\Big[
2s_{_\beta}\Lambda_{_{0,I}}\Lambda_{_{0,J}}({\cal Z}_{_H}^\dagger)_{_{1,\alpha}}
\nonumber\\
&&\hspace{2.2cm}
+\sum\limits_{i=1}^3\Big(\Lambda_{_{0,I}}\Lambda_{_{i,J}}
+\Lambda_{_{i,I}}\Lambda_{_{0,J}}\Big)
({\cal Z}_{_H}^\dagger)_{_{i+1,\alpha}}\Big]
({\cal Z}_{_H})_{_{\alpha,1}}
\Big[s_{_\beta}C_0(x_{_{H_{\alpha}}},x_{_{\tilde h}},x_{_{\tilde A}})
\nonumber\\
&&\hspace{2.2cm}
-c_{_\beta}\Lambda_{_{0,0}}\sqrt{x_{_{\tilde h}}x_{_{\tilde A}}}
e^{-i(\theta_{_\mu}+\theta_{_2})}C(x_{_{H_{\alpha}}},x_{_{\tilde h}}
,x_{_{\tilde A}})\Big]
\nonumber\\
&&\hspace{2.2cm}
+{g_1^2\over |m_{_1}|}e^{-i\theta_{_1}}c_{_\beta}^2\Big[\Lambda_{_{0,I}}
({\bf Y}_{_{L_M}}^\dagger)_{_0}({\bf Y}_{_{R_N}})_{_{0,J}}
+\Lambda_{_{0,J}}({\bf Y}_{_{L_M}}^\dagger)_{_0}({\bf Y}_{_{R_N}})_{_{0,I}}\Big]
({\cal Z}_{_R}^\dagger)_{_{M,i}}({\cal Z}_{_R})_{_{i,N}}
\nonumber\\
&&\hspace{2.2cm}\times
B(x_{_{\tilde h}},x_{_{R_i}})
\nonumber\\
&&\hspace{2.2cm}
+{g_1^2\over |m_{_1}|}e^{-i\theta_{_1}}c_{_\beta}^2\Big[\Lambda_{_{0,I}}
\sum\limits_{j=1}^3({\bf Y}_{_{L_M}}^\dagger)_{_j}({\bf Y}_{_{R_N}})_{_{j,J}}
+\Lambda_{_{0,J}}\sum\limits_{j=1}^3({\bf Y}_{_{L_M}}^\dagger)_{_j}
({\bf Y}_{_{R_N}})_{_{j,I}}\Big]({\cal Z}_{_R}^\dagger)_{_{M,i}}
\nonumber\\
&&\hspace{2.2cm}\times
({\cal Z}_{_R})_{_{i,N}}A(x_{_{R_i}})
\nonumber\\
&&\hspace{2.2cm}
+{g_1^2\over3|m_{_1}|}e^{-i\theta_{_1}}c_{_\beta}^2\Big[\Lambda_{_{0,I}}
({\bf Y}_{_{D_M}}{\bf Y}_{_{S_N}})_{_{0,J}}
+\Lambda_{_{0,J}}({\bf Y}_{_{D_M}}{\bf Y}_{_{S_N}})_{_{0,I}}\Big]
({\cal Z}_{_D}^\dagger)_{_{M,i}}
\nonumber\\
&&\hspace{2.2cm}\times
({\cal Z}_{_D})_{_{i,N}}A(x_{_{D_i}})
\nonumber\\
&&\hspace{2.2cm}
+{g_1^2\over6|m_{_1}|}e^{-i\theta_{_1}}c_{_\beta}^2\Big[
\Lambda_{_{0,I}}({\bf Y}_{_{D_K}}^\dagger)_{_{0,M}}({\bf Y}_{_{S_K}})_{_{N,J}}
+\Lambda_{_{0,J}}({\bf Y}_{_{D_K}}^\dagger)_{_{0,M}}({\bf Y}_{_{S_K}})_{_{N,I}}
\Big]({\cal Z}_{_Q}^\dagger)_{_{M,i}}
\nonumber\\
&&\hspace{2.2cm}\times
({\cal Z}_{_Q})_{_{i,N}}A(x_{_{Q_i}})
\;,\nonumber\\
&&(\delta C_{_2}^{(1)})_{_{J,I}}={g_2^2\mu_{_H}\over2\mu_{_{\rm NP}}^2|m_{_2}|}
e^{-i(\theta_{_\mu}+\theta_{_2})}
c_{_\beta}^2\Big[\Lambda_{_{0,J}}({\bf Y}_{_{R_M}})_{_{0,I}}
+\Lambda_{_{0,I}}({\bf Y}_{_{R_M}})_{_{0,J}}\Big]
({\cal Z}_{_R})_{_{M,i}}({\cal Z}_{_R}^\dagger)_{_{i,N}}
({\bf A}_{_{S_N}})_{_{0,\alpha}}
\nonumber\\
&&\hspace{2.2cm}\times
({\cal Z}_{_H})_{_{\alpha,1}}
C(x_{_{\tilde h}},x_{_{H_{\alpha}}},x_{_{R_i}})
\nonumber\\
&&\hspace{2.2cm}
+{g_2^4\over2|m_{_2}|}e^{-i\theta_{_2}}c_{_\beta}^2\Lambda_{_{0,I}}
\Lambda_{_{0,J}}\Big[s_{_\beta}^2B(x_{_{\tilde h}},x_{_{\tilde A}})
-s_{_\beta}c_{_\beta}\Lambda_{_{0,0}}e^{-i(\theta_{_\mu}+\theta_{_2})}
\sqrt{x_{_{\tilde h}}x_{_{\tilde A}}}P(x_{_{\tilde h}},x_{_{\tilde A}})\Big]
\nonumber\\
&&\hspace{2.2cm}
+{g_1^2g_2^2\over2|m_{_2}|}e^{-i\theta_{_2}}c_{_\beta}^2\Lambda_{_{0,I}}
\Lambda_{_{0,J}}\Big[s_{_\beta}^2B(x_{_{\tilde h}},x_{_{\tilde B}})
-s_{_\beta}c_{_\beta}\Lambda_{_{0,0}}e^{-i(\theta_{_\mu}+\theta_{_1})}
\sqrt{x_{_{\tilde h}}x_{_{\tilde B}}}P(x_{_{\tilde h}},x_{_{\tilde B}})\Big]
\nonumber\\
&&\hspace{2.2cm}
+{g_2^4\over4|m_{_2}|}e^{-i\theta_{_2}}c_{_\beta}^2\Big[
2s_{_\beta}\Lambda_{_{0,I}}\Lambda_{_{0,J}}({\cal Z}_{_H}^\dagger)_{_{1,\alpha}}
+\sum\limits_{i=1}^3\Big(\Lambda_{_{0,I}}\Lambda_{_{i,J}}
+\Lambda_{_{i,I}}\Lambda_{_{0,J}}\Big)
({\cal Z}_{_H}^\dagger)_{_{i+1,\alpha}}\Big]
\nonumber\\
&&\hspace{2.2cm}\times
({\cal Z}_{_H})_{_{\alpha,1}}
\Big[s_{_\beta}C_0(x_{_{H_{\alpha}}},x_{_{\tilde h}},x_{_{\tilde A}})
-c_{_\beta}\Lambda_{_{0,0}}\sqrt{x_{_{\tilde h}}x_{_{\tilde A}}}
e^{-i(\theta_{_\mu}+\theta_{_2})}C(x_{_{H_{\alpha}}},x_{_{\tilde h}},x_{_{\tilde 
A}})\Big]
\nonumber\\
&&\hspace{2.2cm}
+{g_1^2g_2^2\over4|m_{_2}|}e^{-i\theta_{_2}}c_{_\beta}^2\Big[
2s_{_\beta}\Lambda_{_{0,I}}\Lambda_{_{0,J}}({\cal Z}_{_H}^\dagger)_{_{1,\alpha}}
+\sum\limits_{i=1}^3\Big(\Lambda_{_{0,I}}\Lambda_{_{i,J}}
+\Lambda_{_{i,I}}\Lambda_{_{0,J}}\Big)
({\cal Z}_{_H}^\dagger)_{_{i+1,\alpha}}\Big]
\nonumber\\
&&\hspace{2.2cm}\times
({\cal Z}_{_H})_{_{\alpha,1}}
\Big[s_{_\beta}C_0(x_{_{H_{\alpha}}},x_{_{\tilde h}},x_{_{\tilde B}})
-c_{_\beta}\Lambda_{_{0,0}}\sqrt{x_{_{\tilde h}}x_{_{\tilde B}}}
e^{-i(\theta_{_\mu}+\theta_{_1})}C(x_{_{H_{\alpha}}},x_{_{\tilde h}}
,x_{_{\tilde B}})\Big]
\nonumber\\
&&\hspace{2.2cm}
+{g_2^2\over2|m_{_2}|}e^{-i\theta_{_2}}c_{_\beta}^2\Big[
\Lambda_{_{0,I}}({\bf Y}_{_{D_K}}^\dagger)_{_{0,M}}({\bf Y}_{_{S_K}})_{_{N,J}}
+\Lambda_{_{0,J}}({\bf Y}_{_{D_K}}^\dagger)_{_{0,M}}({\bf Y}_{_{S_K}})_{_{N,I}}
\Big]({\cal Z}_{_Q}^\dagger)_{_{M,i}}
\nonumber\\
&&\hspace{2.2cm}\times
({\cal Z}_{_Q})_{_{i,N}}A(x_{_{Q_i}})
\label{aeq1}
\end{eqnarray}
with $x_{_i}=m_{_i}^2/\mu_{_{NP}}^2$.

The corrections from the self-energy diagrams
\begin{eqnarray}
&&(\delta C_{_1}^{(2)})_{_{J,I}}=-{g_1^4\mu_{_H}\over4|m_{_1}|^2}
e^{-i(\theta_{_\mu}+2\theta_{_1})}s_{_\beta}c_{_\beta}^3\Lambda_{_{0,0}}
\Lambda_{_{0,I}}\Lambda_{_{0,J}}A(x_{_{\tilde h}})
\nonumber\\
&&\hspace{2.2cm}
-{g_1^4\mu_{_H}\over4|m_{_1}|^2}e^{-i(\theta_{_\mu}+2\theta_{_1})}c_{_\beta}^3
\Big(s_{_\beta}\Lambda_{_{0,0}}({\cal Z}_{_H}^\dagger)_{_{1,\alpha}}
+\sum\limits_{i=1}^3\Lambda_{_{i,0}}({\cal Z}_{_H}^\dagger)_{_{i,\alpha}}\Big)
({\cal Z}_{_H})_{_{\alpha,1}}\Lambda_{_{0,I}}\Lambda_{_{0,J}}
B(x_{_{\tilde h}},x_{_{H_{\alpha}}})
\nonumber\\
&&\hspace{2.2cm}
-{g_1^4\over4\mu_{_H}}e^{-i(\theta_{_\mu}+2\theta_{_1})}
c_{_\beta}^3\Lambda_{_{0,0}}
\Big\{2s_{_\beta}\Lambda_{_{0,I}}\Lambda_{_{0,J}}A(x_{_{\tilde B}})
+\Big[2s_{_\beta}\Lambda_{_{0,I}}\Lambda_{_{0,J}}({\cal 
Z}_{_H}^\dagger)_{_{1,\alpha}}
+\sum\limits_{i=1}^3\Big(\Lambda_{_{0,I}}\Lambda_{_{i,J}}
\nonumber\\
&&\hspace{2.2cm}
+\Lambda_{_{i,I}}\Lambda_{_{0,J}}\Big)({\cal Z}_{_H}^\dagger)_{_{i,\alpha}}
\Big]({\cal Z}_{_H})_{_{\alpha,1}}B(x_{_{\tilde B}},x_{_{\tilde h}})\Big\}
\nonumber\\
&&\hspace{2.2cm}
-{3g_1^2g_2^2|m_{_2}|\over4\mu_{_H}|m_{_1}|}e^{-i(\theta_{_\mu}+\theta_{_
1}+\theta_{_2})}
c_{_\beta}^3\Lambda_{_{0,0}}
\Big\{2s_{_\beta}\Lambda_{_{0,I}}\Lambda_{_{0,J}}A(x_{_{\tilde A}})
+\Big[2s_{_\beta}\Lambda_{_{0,I}}\Lambda_{_{0,J}}({\cal 
Z}_{_H}^\dagger)_{_{1,\alpha}}
\nonumber\\
&&\hspace{2.2cm}
+\sum\limits_{i=1}^3\Big(\Lambda_{_{i,I}}\Lambda_{_{0,J}}
+\Lambda_{_{0,I}}\Lambda_{_{i,J}}\Big)({\cal Z}_{_H}^\dagger)_{_{i,\alpha}}\Big]
({\cal Z}_{_H})_{_{\alpha,1}}B(x_{_{\tilde A}},x_{_{\tilde h}})\Big\}
\nonumber\\
&&\hspace{2.2cm}
+{3g_1^2g_2^2\over2|m_{_1}|m_{_{H_\alpha}}^2}e^{-i\theta_{_1}}c_{_\beta}
\Big[2s_{_\beta}\Lambda_{_{0,J}}\Lambda_{_{0,I}}({\cal 
Z}_{_H}^\dagger)_{_{1,\alpha}}
+\sum\limits_{i=1}^3\Big(\Lambda_{_{0,I}}\Lambda_{_{i,J}}
+\Lambda_{_{i,I}}\Lambda_{_{0,J}}\Big)({\cal Z}_{_H}^\dagger)_{_{i+1,\alpha}}
\Big]
\nonumber\\
&&\hspace{2.2cm}\times
\Big\{\Big[s_{_\beta}c_{_\beta}\Big(1-|\Lambda_{_{0,0}}|^2\Big)
({\cal Z}_{_H})_{_{\alpha,1}}
-c_{_\beta}\Lambda_{_{0,0}}\sum\limits_{i=1}^3
\Lambda_{_{\alpha,0}}^*({\cal Z}_{_H})_{_{\alpha,i+1}}\Big]
B_0(x_{_{\tilde h}},x_{_{\tilde A}})
\nonumber\\
&&\hspace{2.2cm}
-c_{_\beta}^2\mu_{_H}|m_{_2}|e^{-i(\theta_{_\mu}+\theta_{_2})}\Lambda_{_{0,0}}
({\cal Z}_{_H})_{_{\alpha,1}}B(x_{_{\tilde h}},x_{_{\tilde A}})
-s_{_\beta}\mu_{_H}|m_{_2}|e^{i(\theta_{_\mu}+\theta_{_2})}
\Big[s_{_\beta}\Lambda_{_{0,0}}^*({\cal Z}_{_H})_{_{\alpha,1}}
\nonumber\\
&&\hspace{2.2cm}
+\sum\limits_{i=1}^3\Lambda_{_{i,0}}^*({\cal Z}_{_H})_{_{\alpha,i+1}}\Big]
B(x_{_{\tilde h}},x_{_{\tilde A}})\Big\}
\nonumber\\
&&\hspace{2.2cm}
+{3g_1^2g_2^2\over2|m_{_1}|m_{_{H_\alpha}}^2}e^{-i\theta_{_1}}c_{_\beta}^2
\Big[2s_{_\beta}\Lambda_{_{0,J}}\Lambda_{_{0,I}}({\cal 
Z}_{_H}^\dagger)_{_{1,\alpha}}
+\sum\limits_{i=1}^3\Big(\Lambda_{_{0,I}}\Lambda_{_{i,J}}+\Lambda_{_{i,I}}
\Lambda_{_{0,J}}\Big)({\cal Z}_{_H}^\dagger)_{_{i+1,\alpha}}\Big]
\nonumber\\
&&\hspace{2.2cm}\times
\sum\limits_{j=1}^3\Lambda_{_{0,j}}\Big[s_{_\beta}\Lambda_{_{0,j}}^*
({\cal Z}_{_H})_{_{\alpha,1}}+\sum\limits_{i=1}^3
\Lambda_{_{i,j}}^*({\cal Z}_{_H})_{_{\alpha,i}}\Big]
A_0(x_{_{\tilde A}})
\nonumber\\
&&\hspace{2.2cm}
+{g_1^4\over2|m_{_1}|m_{_{H_\alpha}}^2}e^{-i\theta_{_1}}c_{_\beta}
\Big[2s_{_\beta}\Lambda_{_{0,J}}\Lambda_{_{0,I}}({\cal 
Z}_{_H}^\dagger)_{_{1,\alpha}}
+\sum\limits_{i=1}^3\Big(\Lambda_{_{0,I}}\Lambda_{_{i,J}}
+\Lambda_{_{i,I}}\Lambda_{_{0,J}}\Big)({\cal Z}_{_H}^\dagger)_{_{i+1,\alpha}}
\Big]
\nonumber\\
&&\hspace{2.2cm}\times
\Big\{\Big[s_{_\beta}c_{_\beta}\Big(1-|\Lambda_{_{0,0}}|^2\Big)
({\cal Z}_{_H})_{_{\alpha,1}}
-c_{_\beta}\Lambda_{_{0,0}}\sum\limits_{i=1}^3
\Lambda_{_{\alpha,0}}^*({\cal Z}_{_H})_{_{\alpha,i+1}}\Big]
B_0(x_{_{\tilde h}},x_{_{\tilde B}})
\nonumber\\
&&\hspace{2.2cm}
-c_{_\beta}^2\mu_{_H}|m_{_1}|e^{-i(\theta_{_\mu}+\theta_{_1})}\Lambda_{_{0,0}}
({\cal Z}_{_H})_{_{\alpha,1}}B(x_{_{\tilde h}},x_{_{\tilde B}})
-s_{_\beta}\mu_{_H}|m_{_1}|e^{i(\theta_{_\mu}+\theta_{_1})}
\Big[s_{_\beta}\Lambda_{_{0,0}}^*({\cal Z}_{_H})_{_{\alpha,1}}
\nonumber\\
&&\hspace{2.2cm}
+\sum\limits_{i=1}^3\Lambda_{_{i,0}}^*({\cal Z}_{_H})_{_{\alpha,i+1}}\Big]
B(x_{_{\tilde h}},x_{_{\tilde B}})\Big\}
\nonumber\\
&&\hspace{2.2cm}
+{g_1^4\over2|m_{_1}|m_{_{H_\alpha}}^2}e^{-i\theta_{_1}}c_{_\beta}^2
\Big[2s_{_\beta}\Lambda_{_{0,J}}\Lambda_{_{0,I}}({\cal 
Z}_{_H}^\dagger)_{_{1,\alpha}}
+\sum\limits_{i=1}^3\Big(\Lambda_{_{0,I}}\Lambda_{_{i,J}}+\Lambda_{_{i,I}}
\Lambda_{_{0,J}}\Big)({\cal Z}_{_H}^\dagger)_{_{i+1,\alpha}}\Big]
\nonumber\\
&&\hspace{2.2cm}\times
\sum\limits_{j=1}^3\Lambda_{_{0,j}}\Big[s_{_\beta}\Lambda_{_{0,j}}^*
({\cal Z}_{_H})_{_{\alpha,1}}+\sum\limits_{i=1}^3
\Lambda_{_{i,j}}^*({\cal Z}_{_H})_{_{\alpha,i}}\Big]
A_0(x_{_{\tilde B}})
\;,\nonumber\\
&&(\delta C_{_2}^{(2)})_{_{J,I}}=-{g_2^4\mu_{_H}\over4|m_{_2}|^2}
e^{-i(\theta_{_\mu}+2\theta_{_2})}s_{_\beta}c_{_\beta}^3\Lambda_{_{0,0}}
\Lambda_{_{0,I}}\Lambda_{_{0,J}}A(x_{_{\tilde h}})
-{g_2^4\mu_{_H}\over4|m_{_2}|^2}e^{-i(\theta_{_\mu}+2\theta_{_2})}c_{_\beta}^3
\nonumber\\
&&\hspace{2.2cm}\times
\Big(s_{_\beta}\Lambda_{_{0,0}}({\cal Z}_{_H}^\dagger)_{_{1,\alpha}}
+\sum\limits_{i=1}^3\Lambda_{_{i,0}}({\cal Z}_{_H}^\dagger)_{_{i,\alpha}}
\Big)({\cal Z}_{_H})_{_{\alpha,1}}\Lambda_{_{0,I}}\Lambda_{_{0,J}}
B(x_{_{\tilde h}},x_{_{H_{\alpha}}})
\nonumber\\
&&\hspace{2.2cm}
-{3g_2^4\over4\mu_{_H}}e^{-i(\theta_{_\mu}+2\theta_{_2})}
c_{_\beta}^3\Lambda_{_{0,0}}
\Big\{2s_{_\beta}\Lambda_{_{0,I}}\Lambda_{_{0,J}}A(x_{_{\tilde A}})
+\Big[2s_{_\beta}\Lambda_{_{0,I}}\Lambda_{_{0,J}}({\cal 
Z}_{_H}^\dagger)_{_{1,\alpha}}
+\sum\limits_{i=1}^3\Big(\Lambda_{_{0,I}}\Lambda_{_{i,J}}
\nonumber\\
&&\hspace{2.2cm}
+\Lambda_{_{i,I}}\Lambda_{_{0,J}}\Big)({\cal Z}_{_H}^\dagger)_{_{i,\alpha}}
\Big]({\cal Z}_{_H})_{_{\alpha,1}}B(x_{_{\tilde A}},x_{_{\tilde h}})\Big\}
-{g_1^2g_2^2|m_{_1}|\over4\mu_{_H}|m_{_2}|}e^{-i(\theta_{_\mu}+\theta_{_1
}+\theta_{_2})}
c_{_\beta}^3\Lambda_{_{0,0}}
\nonumber\\
&&\hspace{2.2cm}\times
\Big\{2s_{_\beta}\Lambda_{_{0,I}}\Lambda_{_{0,J}}A(x_{_{\tilde B}})
+\Big[2s_{_\beta}\Lambda_{_{0,I}}\Lambda_{_{0,J}}({\cal 
Z}_{_H}^\dagger)_{_{1,\alpha}}
+\sum\limits_{i=1}^3\Big(\Lambda_{_{i,I}}\Lambda_{_{0,J}}
+\Lambda_{_{0,I}}\Lambda_{_{i,J}}\Big)({\cal Z}_{_H}^\dagger)_{_{i,\alpha}}\Big]
\nonumber\\
&&\hspace{2.2cm}\times
({\cal Z}_{_H})_{_{\alpha,1}}B(x_{_{\tilde B}},x_{_{\tilde h}})\Big\}
\nonumber\\
&&\hspace{2.2cm}
+{3g_2^4\over2|m_{_2}|m_{_{H_\alpha}}^2}e^{-i\theta_{_2}}c_{_\beta}
\Big[2s_{_\beta}\Lambda_{_{0,J}}\Lambda_{_{0,I}}({\cal 
Z}_{_H}^\dagger)_{_{1,\alpha}}
+\sum\limits_{i=1}^3\Big(\Lambda_{_{0,I}}\Lambda_{_{i,J}}
+\Lambda_{_{i,I}}\Lambda_{_{0,J}}\Big)({\cal Z}_{_H}^\dagger)_{_{i+1,\alpha}}
\Big]
\nonumber\\
&&\hspace{2.2cm}\times
\Big\{\Big[s_{_\beta}c_{_\beta}\Big(1-|\Lambda_{_{0,0}}|^2\Big)
({\cal Z}_{_H})_{_{\alpha,1}}
-c_{_\beta}\Lambda_{_{0,0}}\sum\limits_{i=1}^3
\Lambda_{_{\alpha,0}}^*({\cal Z}_{_H})_{_{\alpha,i+1}}\Big]
B_0(x_{_{\tilde h}},x_{_{\tilde A}})
\nonumber\\
&&\hspace{2.2cm}
-c_{_\beta}^2\mu_{_H}|m_{_2}|e^{-i(\theta_{_\mu}+\theta_{_2})}\Lambda_{_{0,0}}
({\cal Z}_{_H})_{_{\alpha,1}}B(x_{_{\tilde h}},x_{_{\tilde A}})
-s_{_\beta}\mu_{_H}|m_{_2}|e^{i(\theta_{_\mu}+\theta_{_2})}
\Big[s_{_\beta}\Lambda_{_{0,0}}^*({\cal Z}_{_H})_{_{\alpha,1}}
\nonumber\\
&&\hspace{2.2cm}
+\sum\limits_{i=1}^3\Lambda_{_{i,0}}^*({\cal Z}_{_H})_{_{\alpha,i+1}}\Big]
B(x_{_{\tilde h}},x_{_{\tilde A}})\Big\}
\nonumber\\
&&\hspace{2.2cm}
+{3g_2^4\over2|m_{_2}|m_{_{H_\alpha}}^2}e^{-i\theta_{_2}}c_{_\beta}^2
\Big[2s_{_\beta}\Lambda_{_{0,J}}\Lambda_{_{0,I}}({\cal 
Z}_{_H}^\dagger)_{_{1,\alpha}}
+\sum\limits_{i=1}^3\Big(\Lambda_{_{0,I}}\Lambda_{_{i,J}}+\Lambda_{_{i,I}}
\Lambda_{_{0,J}}\Big)({\cal Z}_{_H}^\dagger)_{_{i+1,\alpha}}\Big]
\nonumber\\
&&\hspace{2.2cm}\times
\sum\limits_{j=1}^3\Lambda_{_{0,j}}\Big[s_{_\beta}\Lambda_{_{0,j}}^*
({\cal Z}_{_H})_{_{\alpha,1}}+\sum\limits_{i=1}^3
\Lambda_{_{i,j}}^*({\cal Z}_{_H})_{_{\alpha,i}}\Big]
A_0(x_{_{\tilde A}})
\nonumber\\
&&\hspace{2.2cm}
+{g_1^2g_2^2\over2|m_{_2}|m_{_{H_\alpha}}^2}e^{-i\theta_{_2}}c_{_\beta}
\Big[2s_{_\beta}\Lambda_{_{0,J}}\Lambda_{_{0,I}}({\cal 
Z}_{_H}^\dagger)_{_{1,\alpha}}
+\sum\limits_{i=1}^3\Big(\Lambda_{_{0,I}}\Lambda_{_{i,J}}
+\Lambda_{_{i,I}}\Lambda_{_{0,J}}\Big)({\cal Z}_{_H}^\dagger)_{_{i+1,\alpha}}
\Big]
\nonumber\\
&&\hspace{2.2cm}\times
\Big\{\Big[s_{_\beta}c_{_\beta}\Big(1-|\Lambda_{_{0,0}}|^2\Big)
({\cal Z}_{_H})_{_{\alpha,1}}
-c_{_\beta}\Lambda_{_{0,0}}\sum\limits_{i=1}^3
\Lambda_{_{\alpha,0}}^*({\cal Z}_{_H})_{_{\alpha,i+1}}\Big]
B_0(x_{_{\tilde h}},x_{_{\tilde B}})
\nonumber\\
&&\hspace{2.2cm}
-c_{_\beta}^2\mu_{_H}|m_{_1}|e^{-i(\theta_{_\mu}+\theta_{_1})}\Lambda_{_{0,0}}
({\cal Z}_{_H})_{_{\alpha,1}}B(x_{_{\tilde h}},x_{_{\tilde B}})
-s_{_\beta}\mu_{_H}|m_{_1}|e^{i(\theta_{_\mu}+\theta_{_1})}
\Big[s_{_\beta}\Lambda_{_{0,0}}^*({\cal Z}_{_H})_{_{\alpha,1}}
\nonumber\\
&&\hspace{2.2cm}
+\sum\limits_{i=1}^3\Lambda_{_{i,0}}^*({\cal Z}_{_H})_{_{\alpha,i+1}}\Big]
B(x_{_{\tilde h}},x_{_{\tilde B}})\Big\}
\nonumber\\
&&\hspace{2.2cm}
+{g_1^2g_2^2\over2|m_{_2}|m_{_{H_\alpha}}^2}e^{-i\theta_{_2}}c_{_\beta}^2
\Big[2s_{_\beta}\Lambda_{_{0,J}}\Lambda_{_{0,I}}({\cal 
Z}_{_H}^\dagger)_{_{1,\alpha}}
+\sum\limits_{i=1}^3\Big(\Lambda_{_{0,I}}\Lambda_{_{i,J}}+\Lambda_{_{i,I}}
\Lambda_{_{0,J}}\Big)({\cal Z}_{_H}^\dagger)_{_{i+1,\alpha}}\Big]
\nonumber\\
&&\hspace{2.2cm}\times
\sum\limits_{j=1}^3\Lambda_{_{0,j}}\Big[s_{_\beta}\Lambda_{_{0,j}}^*
({\cal Z}_{_H})_{_{\alpha,1}}+\sum\limits_{i=1}^3
\Lambda_{_{i,j}}^*({\cal Z}_{_H})_{_{\alpha,i}}\Big]
A_0(x_{_{\tilde B}})
\;,\nonumber\\
\label{aeq2}
\end{eqnarray}
where $x_{_i}=m_{_i}^2/\mu_{_{\rm NP}}^2$, and the nonzero Yukawa couplings are
\begin{eqnarray}
&&({\bf Y}_{_{R_K}})_{_{0,1}}=h^e_{_{1,K}}{\mu_{_{H_1}}\over\mu_{_H}}
+{\Big(h^e_{_{2,K}}|\epsilon_{_1}|e^{i\varphi_{_1}}-2\lambda_{_{12K}}|\epsilon_{_0}|\Big)
|\epsilon_{_2}|e^{-i\varphi_{_2}}\over\mu_{_H}\mu_{_{H_1}}}
\nonumber\\
&&\hspace{2.2cm}
+{\Big(h^e_{_{3,K}}|\epsilon_{_1}|e^{i\varphi_{_1}}-2\lambda_{_{13K}}|\epsilon_{_0}|\Big)
|\epsilon_{_3}|e^{-i\varphi_{_3}}\over\mu_{_H}\mu_{_{H_1}}}
\;,\nonumber\\
&&({\bf Y}_{_{R_K}})_{_{0,2}}={\mu_{_{H_2}}\Big(h^e_{_{2,K}}|\epsilon_{_0}|
-2\lambda_{_{21K}}|\epsilon_{_1}|e^{-i\varphi_{_1}}\Big)\over\mu_{_{H_1}}\mu_{_H}}
+{|\epsilon_{_2}\epsilon_{_3}|e^{i(\varphi_{_2}-\varphi_{_3})}\Big(h^e_{_
{3,K}}|\epsilon_{_0}|
+2\lambda_{_{13K}}|\epsilon_{_1}|e^{-i\varphi_{_1}}\Big)
\over\mu_{_{H_1}}\mu_{_{H_2}}\mu_{_H}}
\nonumber\\
&&\hspace{2.2cm}
-2\lambda_{_{23K}}{\mu_{_{H_1}}|\epsilon_{_3}|\over\mu_{_{H_2}}\mu_{_H}}
e^{-i\varphi_{_3}}
\;,\nonumber\\
&&({\bf Y}_{_{R_K}})_{_{0,3}}={1\over\mu_{_{H_2}}}\Big\{h^e_{_{3,K}}
|\epsilon_{_0}|-2\Big(\lambda_{_{31K}}|\epsilon_{_1}|e^{-i\varphi_{_1}}
+\lambda_{_{32K}}|\epsilon_{_2}|e^{-i\varphi_{_2}}\Big)\Big\}
\;,\nonumber\\
&&({\bf Y}_{_{R_K}})_{_{1,2}}={h^e_{_{1,K}}|\epsilon_{_2}|e^{i\varphi_{_2}}
-h^e_{_{2,K}}|\epsilon_{_1}|e^{i\varphi_{_1}}\over\mu_{_{H_2}}}
+2\lambda_{_{12K}}{|\epsilon_{_0}|\over\mu_{_{H_2}}}
\;,\nonumber\\
&&({\bf Y}_{_{R_K}})_{_{1,3}}=h^e_{_{1,K}}{\mu_{_{H_1}}|\epsilon_{_3}|
\over\mu_{_{H_2}}\mu_{_H}}e^{i\varphi_{_3}}
+{|\epsilon_{_2}\epsilon_{_3}|e^{i(\varphi_{_3}-\varphi_{_2})}
\Big(h^e_{_{2,K}}|\epsilon_{_1}|e^{-i\varphi_{_1}}
-2\lambda_{_{12K}}|\epsilon_{_0}|\Big)\over\mu_{_{H_1}}\mu_{_{H_2}}\mu_{_
H}}
\nonumber\\
&&\hspace{2.2cm}
-{\mu_{_{H_2}}\Big(h^e_{_{3,K}}|\epsilon_{_1}|e^{-i\varphi_{_1}}
-2\lambda_{_{13K}}|\epsilon_{_0}|\Big)\over\mu_{_{H_1}}\mu_{_H}}
\;,\nonumber\\
&&({\bf Y}_{_{R_K}})_{_{2,3}}=2\lambda_{_{23K}}{\mu_{_{H_1}}\over\mu_{_H}}
+{|\epsilon_{_3}|e^{i\varphi_{_3}}\Big(h^e_{_{2,K}}|\epsilon_{_0}|
+2\lambda_{_{12K}}|\epsilon_{_1}|e^{-i\varphi_{_1}}\Big)\over\mu_{_{H_1}}\mu_{_H}}
\nonumber\\
&&\hspace{2.2cm}
-{|\epsilon_{_2}|e^{i\varphi_{_2}}\Big(h^e_{_{3,K}}|\epsilon_{_0}|
+2\lambda_{_{13K}}|\epsilon_{_1}|e^{-i\varphi_{_1}}\Big)\over\mu_{_{H_1}}\mu_{_H}}
\;,\nonumber\\
&&({\bf Y}_{_{L_K}})_{_0}=\sum\limits_{I=1}^3{\Big(\upsilon_{_0}|\epsilon_{_I}|
e^{-i\varphi_{_I}}-\upsilon_{_I}|\epsilon_{_0}|\Big)h^e_{_{I,K}}\over
\upsilon_{_d}\mu_{_H}}
+\sum\limits_{I,J}^3{\Big(\upsilon_{_J}|\epsilon_{_I}|
e^{-i\varphi_{_I}}-\upsilon_{_I}|\epsilon_{_J}|e^{-i\varphi_{_J}}\Big)\over
\upsilon_{_d}\mu_{_H}}\lambda_{_{IJK}}
\;,\nonumber\\
&&({\bf Y}_{_{L_K}})_{_1}={|\epsilon_{_0}|
{\cal H}^e_{_{1,K}}+\sum\limits_{I=1}^3\upsilon_{_I}h^e_{_{I,K}}
|\epsilon_{_1}|e^{i\varphi_{_1}}\over\upsilon_{_d}\mu_{_{H_1}}}
\;,\nonumber\\
&&({\bf Y}_{_{L_K}})_{_2}={1\over\upsilon_{_d}\mu_{_{H_1}}\mu_{_{H_2}}}
\Big\{\mu_{_{H_1}}^2{\cal H}^e_{_{2,K}}
-{\cal H}^e_{_{1,K}}|\epsilon_{_1}\epsilon_{_2}|
e^{i(\varphi_{_2}-\varphi_{_1})}
+\sum\limits_{I=1}^3\upsilon_{_I}h^e_{_{I,K}}|\epsilon_{_0}\epsilon_{_2}|
e^{i\varphi_{_2}}\Big\}
\;,\nonumber\\
&&({\bf Y}_{_{L_K}})_{_3}={1\over\upsilon_{_d}\mu_{_{H_2}}\mu_{_H}}
\Big\{\mu_{_{H_2}}^2{\cal H}^e_{_{3,K}}-{\cal 
H}^e_{_{1,K}}|\epsilon_{_1}\epsilon_{_3}|
e^{i(\varphi_{_3}-\varphi_{_1})}-{\cal H}^e_{_{2,K}}|\epsilon_{_2}\epsilon_{_3}|
e^{i(\varphi_{_3}-\varphi_{_1})}
\nonumber\\
&&\hspace{2.2cm}
+\sum\limits_{I=1}^3\upsilon_{_I}h^e_{_{I,K}}|\epsilon_{_0}\epsilon_{_3}|
e^{i\varphi_{_3}}\Big\}
\;,\nonumber\\
&&({\bf Y}_{_{D_K}})_{_{0,I}}={\upsilon_{_0}h^d_{_{I,K}}+\sum\limits_{J=1}^3
\upsilon_{_J}\lambda^\prime_{_{JIK}}\over\upsilon_{_d}}
\;,\nonumber\\
&&({\bf Y}_{_{S_K}})_{_{I,1}}={h^d_{_{I,K}}|\epsilon_{_1}|e^{i\varphi_{_1}}
-\lambda^\prime_{_{1IK}}|\epsilon_{_0}|\over\mu_{_{H_1}}}
\;,\nonumber\\
&&({\bf Y}_{_{S_K}})_{_{I,2}}={h^d_{_{I,K}}|\epsilon_{_0}\epsilon_{_2}|
e^{i\varphi_{_2}}+\lambda^\prime_{_{1IK}}|\epsilon_{_1}\epsilon_{_2}|
e^{i(\varphi_{_2}-\varphi_{_1})}
\over\mu_{_{H_1}}\mu_{_{H_2}}}
-{\mu_{_{H_1}}\lambda^\prime_{_{2IK}}\over\mu_{_{H_2}}}
\;,\nonumber\\
&&({\bf Y}_{_{S_K}})_{_{I,3}}={1\over\mu_{_{H_2}}\mu_{_{H}}}\Big\{
h^d_{_{I,K}}|\epsilon_{_0}\epsilon_{_3}e^{i\varphi_{_3}}
+\lambda^\prime_{_{1IK}}|\epsilon_{_1}\epsilon_{_3}|
e^{i(\varphi_{_3}-\varphi_{_1})}
+\lambda^\prime_{_{2IK}}|\epsilon_{_2}\epsilon_{_3}|
e^{i(\varphi_{_3}-\varphi_{_2})}
\nonumber\\
&&\hspace{2.2cm}
-\mu_{_{H_2}}^2\lambda^\prime_{_{3IK}}\Big\}
\label{aeq3}
\end{eqnarray}
with 
\begin{eqnarray}
&&{\cal H}^e_{_{I,J}}=\upsilon_{_0}h^e_{_{I,J}}+2\sum\limits_{\rho\neq I}
\upsilon_{_\rho}\lambda_{_{\rho IJ}}\;.
\label{aeq4}
\end{eqnarray}
The couplings between the Higgs and right-handed slepton are
given by
\begin{eqnarray}
&&({\bf A}_{_{S_I}})_{_{0,\alpha}}=s_{_\beta}^2\sum\limits_{\rho=1}^3
{\upsilon_{_\rho}\over\upsilon_{_d}}\eta_{\rho}^I({\cal Z}_{_H})_{_{\alpha,0}}
+\sum\limits_{K=1}^3\Big[s_{_\beta}{\upsilon_{_{d_{K-1}}}^2\eta_K^I
-\upsilon_{_K}\sum\limits_{\beta=0}^{K-1}\upsilon_{_\beta}\eta_\alpha^I
\over\upsilon_{_{d_{K-1}}}\upsilon_{_{d_K}}}
+c_{_\beta}\xi_K^I\Big]({\cal Z}_{_H})_{_{\alpha,K}}
\label{aeq5}
\end{eqnarray}
with
\begin{eqnarray}
&&\eta_{0}^I=\sum\limits_{K=1}^3\epsilon_{_K}^*h_{_{K,I}}^e,\;
\nonumber\\
&&\eta_{K}^I=-\epsilon_{_0}^*h_{_{K,I}}^e-2\sum\limits_{J=1}^3
\epsilon_{_J}^*\lambda_{_{JKI}},\;
\nonumber\\
&&\xi_1^I={\upsilon_{_{d_1}}\over\upsilon_{_d}}A_{_{1I}}^e
+{\upsilon_{_2}(\upsilon_{_1}A_{_{2I}}^e-2\upsilon_{_0}A_{_{12I}})
\over\upsilon_{_{d_1}}\upsilon_{_d}}
+{\upsilon_{_3}(\upsilon_{_1}A_{_{3I}}^e-2\upsilon_{_0}A_{_{13I}})
\over\upsilon_{_{d_1}}\upsilon_{_d}},\;
\nonumber\\
&&\xi_2^I={\upsilon_{_{d_2}}(\upsilon_{_0}A_{_{2I}}^e+2\upsilon_{_1}
A_{_{12I}})\over\upsilon_{_{d_1}}\upsilon_{_d}}
-2{\upsilon_{_3}\upsilon_{_{d_1}}\over\upsilon_{_{d_2}}\upsilon_{_d}}
A_{_{23I}}+{\upsilon_{_0}\upsilon_{_2}\upsilon_{_3}\over\upsilon_{_{d_1}}
\upsilon_{_{d_2}}\upsilon_{_d}}A_{_{3I}}^e
+2{\upsilon_{_1}\upsilon_{_2}
\upsilon_{_3}\over\upsilon_{_{d_1}}\upsilon_{_{d_2}}\upsilon_{_d}}
A_{_{13I}},\;
\nonumber\\
&&\xi_3^I={\upsilon_{_0}A_{_{3I}}^e+2\upsilon_{_1}A_{_{13I}}
+2\upsilon_{_2}A_{_{23I}}\over\upsilon_{_{d_2}}}\;.
\label{aeq6}
\end{eqnarray}
\end{widetext}

The loop-integral functions are
\begin{eqnarray}
&&A(x)={1-\ln x\over(4\pi)^2}\;,\nonumber\\
&&A_0(x)=x{1-\ln x\over(4\pi)^2}\;,\nonumber\\
&&B(x,y)={1\over(4\pi)^2}\Big[1+{x\ln x\over y-x}+{y\ln y\over 
x-y}\Big]\;,\nonumber\\
&&B_0(x,y)={1\over(4\pi)^2}\Big[x+y+{x^2\ln x\over y-x}+{y^2\ln y\over 
x-y}\Big]\;,\nonumber\\
&&P(x,y)={1\over(4\pi)^2}\Big[{\ln x\over x-y}+{\ln y\over y-x}\Big]\;,\nonumber\\
&&C(x,y,z)={1\over(4\pi)^2}\Big[{x\ln x\over(y-x)(z-x)}
\nonumber\\&&\hspace{0.6cm}
+{y\ln y\over(x-y)(z-y)}
+{z\ln z\over(x-z)(y-z)}\Big]\;,\nonumber\\
&&C_0(x,y,z)={1\over(4\pi)^2}\Big[{x^2\ln x\over(y-x)(z-x)}
\nonumber\\&&\hspace{0.6cm}
+{y^2\ln y\over(x-y)(z-y)}
+{z^2\ln z\over(x-z)(y-z)}-1\Big]\;.
\label{aeq7}
\end{eqnarray}


\begin{thebibliography}{}
\bibitem{exp}S.~Fukuda {\it et al.}  [Super-Kamiokande Collaboration],
Phys.\ Lett.\ B {\bf 539}, 179(2002)[arXiv:hep-ex/0205075];
Q.~R.~Ahmad {\it et al.}  [SNO Collaboration],
Phys.\ Rev.\ Lett.\  {\bf 87}, 071301(2001) [arXiv:nucl-ex/0106015];
Q.~R.~Ahmad {\it et al.}  [SNO Collaboration],
Phys.\ Rev.\ Lett.\  {\bf 89}, 011301(2002)[arXiv:nucl-ex/0204008];
Q.~R.~Ahmad {\it et al.}  [SNO Collaboration],
Phys.\ Rev.\ Lett.\  {\bf 89}, 011302 (2002)
[arXiv:nucl-ex/0204009];
S.~N.~Ahmed {\it et al.}  [SNO Collaboration],
arXiv:nucl-ex/0309004; K.~Eguchi {\it et al.}  [KamLAND Collaboration],
Phys.\ Rev.\ Lett.\  {\bf 90}, 021802 (2003)
[arXiv:hep-ex/0212021]; T. Araki {\it et al.}  [KamLAND Collaboration],
arXiv:hep-ex/0406035;
Y.~Ashie {\it et al.}  [Super-Kamiokande Collaboration],
arXiv:hep-ex/0404034.
\bibitem{mssm}H. P. Nilles, Phys. Rept. {\bf 110}, 1(1984);
H. E. Haber, G. L. Kane, Phys. Rept. {\bf 117}, 75(1985). 
\bibitem{Rp} G.R.~Farrar and P.~Fayet, Nucl.\ Phys.\ {\bf B76} (1978)575.
\bibitem{rpv}M.~Chemtob, hep-ph/0406029; R.~Barbier {\it et al.},
hep-ph/0406039;  M.~A.~Diaz, J.~C.~Romao, J.~W.~F.~Valle,
Nucl. Phys. B {\bf 524}, 23(1998); M.~A.~Diaz, J.~Ferrandis, 
J.~C.~Romao, J.~W.~F.~Valle, Phys. Lett. B {\bf 453}, 263(1999);
M.~B.~Magro, F.~de~Campos, O.~J.~P.~Eboli, W.~Porod , D.~Restrepo, J.~W.~F.~Valle, 
JHEP 
{\bf 0309}, 071(2003);  A.~Bartl, M.~Hirsch, T.~Kernreiter, W.~Porod, 
J.~W.~F.~Valle, JHEP {\bf 0311},005(2003); M.~Hirsch, J.~W.~F.~Valle, 
hep-ph/0405015;
B.~Dutta, C.~S.~Kim, S.~Oh, Phys.\ Lett.\ B {\bf 535}, 249(2002); 
D.~A.~Restrepo Quintero, hep-ph/0111198;
F.~De Campos, M.~A.~Diaz, O.~J.~P.~Eboli, M.~B.~Magro, P.~G.~Mercadante,
Nucl.\ Phys.\ B {\bf 623}, 47(2002); R.~Adhikari, E.~Ma, G.~Rajasekaran,
Phys.\ Rev.\ D {\bf 65}, 077703(2002);  C.~H.~Chang, T.~F.~Feng, Eur.\ Phys.\ J.\
C. {\bf 12}, 137(2000); M.~Frank, K.~Huitu, Phys.\
Rev.\ D {\bf 64}, 095015(2001); W.~Porod, M.~Hirsch, J.~Romao, J.~W.~F.~Valle,
Phys.\ Rev.\ D {\bf 63}, 115004(2001);  A.~Datta, P.~Konar, B.~Mukhopadhyaya,
Phys.\ Rev.\ D {\bf 63}, 095009(2001);  T.~Hambye, E.~Ma, U.~Sarkar,
Nucl.\ Phys.\ B {\bf 590}, 429(2000); A.~Datta, B.~Mukhopadhyaya, Phys.\ Rev.\ 
Lett.\
{\bf 85}, 248(2000);  
A.~Datta, B.~Mukhopadhyaya, F.~Vissani, Phys.\ Lett.\ B {\bf 492}, 324(2000);
J.~L.~Chkareuli, I.~G.~Gogoladze, A.~B.~Kobakhidze, M.~G.~Green, D.~E.~Hutchcroft,
Phys.\ Rev.\ D {\bf 62}, 015014(2000); B.~Mukhopadhyaya, Pramana {\bf 54}, 
147(2000);
B.~Mukhopadhyaya, S.~Roy, Phys.\ Rev.\ D {\bf 60}, 115012(1999).
\bibitem{feng1}C.~S.~Aulakh and R.~N.~Mohapatra,
Phys.\ Lett.\ B {\bf 119}, 136 (1982);
L.~J.~Hall and M.~Suzuki, Nucl.\ Phys.\ B {\bf 231}, 419 (1984);
I.~H.~Lee, Phys.\ Lett.\ B {\bf 138}, 121 (1984);
Nucl.\ Phys.\ B {\bf 246}, 120 (1984);
G.~G.~Ross and J.~W.~Valle, Phys.\ Lett.\ B {\bf 151}, 375 (1985);
J.~R.~Ellis, G.~Gelmini, C.~Jarlskog, G.~G.~Ross and J.~W.~Valle,
Phys.\ Lett.\ B {\bf 150}, 142 (1985);
S.~Dawson, Nucl.\ Phys.\ B {\bf 261}, 297 (1985);
A.~Santamaria and J.~W.~Valle, Phys.\ Lett.\ B {\bf 195}, 423 (1987);
K.~S.~Babu and R.~N.~Mohapatra, Phys.\ Rev.\ Lett.\  {\bf 64}, 1705 (1990);
R.~Barbieri, M.~M.~Guzzo, A.~Masiero and D.~Tommasini,
Phys.\ Lett.\ B {\bf 252}, 251 (1990);
E.~Roulet and D.~Tommasini, Phys.\ Lett.\ B {\bf 256}, 218 (1991);
K.~Enqvist, A.~Masiero and A.~Riotto,
Nucl.\ Phys.\ B {\bf 373}, 95 (1992); J.~C.~Romao and J.~W.~Valle,
Nucl.\ Phys.\ B {\bf 381}, 87 (1992);
R.~M.~Godbole, P.~Roy and X.~Tata, Nucl.\ Phys.\ B {\bf 401}, 67 (1993)
[hep-ph/9209251];
A.~S.~Joshipura and M.~Nowakowski, Phys.\ Rev.\ D {\bf 51}, 2421 (1995)
[hep-ph/9408224]; Phys.\ Rev.\ D {\bf 51}, 5271 (1995) [hep-ph/9403349];
M.~Nowakowski and A.~Pilaftsis, Nucl.\ Phys.\ B {\bf 461}, 19 (1996)
[hep-ph/9508271]; F.~M.~Borzumati, Y.~Grossman, E.~Nardi and Y.~Nir,
Phys.\ Lett.\ B {\bf 384}, 123 (1996) [hep-ph/9606251];
S.~Roy and B.~Mukhopadhyaya, Phys.\ Rev.\ D {\bf 55}, 7020 (1997)
[hep-ph/9612447]; A.~S.~Joshipura, V.~Ravindran and S.~K.~Vempati,
Phys.\ Lett.\ B {\bf 451}, 98 (1999) [hep-ph/9706482];
M.~Drees, S.~Pakvasa, X.~Tata and T.~ter Veldhuis, Phys.\ Rev.\ D {\bf 57}, 5335 
(1998)
[hep-ph/9712392]; R.~Adhikari and G.~Omanovic,
Phys.\ Rev.\ D {\bf 59}, 073003 (1999); A.~S.~Joshipura and S.~K.~Vempati,
Phys.\ Rev.\ D {\bf 60}, 095009 (1999) [hep-ph/9808232].
B.~Mukhopadhyaya, S.~Roy and F.~Vissani, Phys.\ Lett.\ B {\bf 443}, 191 (1998)
[hep-ph/9808265];
K.~Choi, K.~Hwang and E.~J.~Chun, Phys.\ Rev.\ D {\bf 60}, 031301 (1999)
[hep-ph/9811363];
S.~Rakshit, G.~Bhattacharyya and A.~Raychaudhuri,
Phys.\ Rev.\ D {\bf 59}, 091701 (1999) [hep-ph/9811500];
D.~E.~Kaplan and A.~E.~Nelson, JHEP {\bf 0001}, 033 (2000)
[hep-ph/9901254];
A.~S.~Joshipura and S.~K.~Vempati, Phys.\ Rev.\ D {\bf 60}, 111303 (1999)
[hep-ph/9903435];
S.~Y.~Choi, E.~J.~Chun, S.~K.~Kang and J.~S.~Lee,
Phys.\ Rev.\ D {\bf 60}, 075002 (1999) [hep-ph/9903465];
A.~Datta, B.~Mukhopadhyaya and S.~Roy, Phys.\ Rev.\ D {\bf 61}, 055006 (2000)
[hep-ph/9905549];
G.~Bhattacharyya, H.~V.~Klapdor-Kleingrothaus and H.~Paes,
Phys.\ Lett.\ B {\bf 463}, 77 (1999) [hep-ph/9907432];
A.~Abada and M.~Losada, Nucl.\ Phys.\ B {\bf 585}, 45 (2000)
[hep-ph/9908352];
O.~Haug, J.~D.~Vergados, A.~Faessler and S.~Kovalenko,
Nucl.\ Phys.\ B {\bf 565}, 38 (2000) [hep-ph/9909318];
E.~J.~Chun and S.~K.~Kang, Phys.\ Rev.\ D {\bf 61}, 075012 (2000)
[hep-ph/9909429];
F.~Takayama and M.~Yamaguchi, Phys.\ Lett.\ B {\bf 476}, 116 (2000)
[hep-ph/9910320];
R.~Kitano and K.~y.~Oda, Phys.\ Rev.\ D {\bf 61}, 113001 (2000)
[hep-ph/9911327];
M.~Hirsch, M.~A.~Diaz, W.~Porod, J.~C.~Romao and J.~W.~Valle,
Phys.\ Rev.\ D {\bf 62}, 113008 (2000)
[Erratum-ibid.\ D {\bf 65}, 119901 (2002)] [hep-ph/0004115];
A.~S.~Joshipura, R.~D.~Vaidya and S.~K.~Vempati,
Phys.\ Rev.\ D {\bf 62}, 093020 (2000) [hep-ph/0006138].
J.~M.~Mira, E.~Nardi, D.~A.~Restrepo and J.~W.~Valle,
Phys.\ Lett.\ B {\bf 492}, 81 (2000) [hep-ph/0007266];
T.~F.~Feng and X.~Q.~Li, Phys.\ Rev.\ D {\bf 63}, 073006 (2001)
[hep-ph/0012300];
A.~S.~Joshipura, R.~D.~Vaidya and S.~K.~Vempati,
Phys.\ Rev.\ D {\bf 65}, 053018 (2002) [hep-ph/0107204];
V.~D.~Barger, T.~Han, S.~Hesselbach and D.~Marfatia,
Phys.\ Lett.\ B {\bf 538}, 346 (2002) [hep-ph/0108261];
A.~S.~Joshipura, R.~D.~Vaidya and S.~K.~Vempati,
Nucl.\ Phys.\ B {\bf 639}, 290 (2002) [hep-ph/0203182];
S.~K.~Vempati, hep-ph/0203219;
M.~A.~Diaz, M.~Hirsch, W.~Porod, J.~C.~Romao and J.~W.~F.~Valle,
Phys.\ Rev.\ D {\bf 68}, 013009 (2003) [hep-ph/0302021];
A. Abada, G. Bhattacharyya, M. Losada, Phys. Rev. D 
{\bf 66}, 071701(2002); M. A. Diaz, R. A. Lineros, M. A. Rivera
Phys. Rev. D {\bf 67}, 115004(2003);  M.A. Diaz, M. Hirsch, W. Porod, 
J.C. Romao, J.W.F. Valle, Phys. Rev. D {\bf 68}, 013009(2003);
D. A. Sierra, M. Hirsch, J.W.F. Valle, A. Villanova del Moral,
Phys. Rev. D {\bf 68}, 033006(2003); S. P. Das, A. Datta, 
M. Guchait, hep-ph/0309168;  Y. Grossman, S. Rakshit,
Phys. Rev. D {\bf 69}, 093002(2004); S. Rakshit, hep-ph/0406168.
\bibitem{Burgess}C. J. C. Burgess, H. J. Schnitzer, Nucl. Phys. {\bf B228},
464(1983); C. N. Leung, S. T. Love, S. Rao, Z. Phys. {\bf C31}, 433(1986);
B. Buchmuller, D. Wyler, Nucl. Phys. {\bf B268}, 621(1986);
K. Hagiwara, S. Ishihara, R. Szalapski, D. Zeppenfeld,
Phys. Rev.{\bf D48}, 2182(1993); B. Grzadkowski, Z. Hioki, K. Ohkuma,
J. Wudka, hep-ph/0310159; V. Barger, T. Han, P. Langacker, B. McElrath, P. Zerwas,
Phys. Rev. {\bf D67}, 115001(2003).
\bibitem{Weinberg}S. Weinberg, Phys. Rev. Lett. {\bf 43}, 1566(1979);
F. Wilczek, A. Zee, {\it ibid.} {\bf 43}, 1571(1979).
\bibitem{Broncano}A. Broncano, M. B. Gavela and E. Jenkins, Nucl.
Phys. B {\bf 672}, 163(2003).
\bibitem{feng2}T.-F.~Feng, X.-Q.~Li, J. ~Maamalpi, Phys. Rev. D.
{\bf 69}, 115007(2004).
\bibitem{Hempfling}R.~Hempfling, Nucl. Phys. B. {\bf 478}, 3(1996);
H.~P.~Nilles, and N.~Polonsky, {\it ibid.} {\bf 484}, 33(1997).
\bibitem{Gonzalez-Garcia}M. C. Gonzalez-Garcia, Y. Nir, Rev. Mod.
Phys. {\bf 75}, 345(2003); O. Peres, A. Smirnov, Nucl.Phys. B
{\bf 680}, 479(2004); G. Fogli, E. Marrone, D. Montanino, A. Palazzo, 
A. Rotunno, Phys. Rev. D {\bf 69}, 017301(2004); M.C. Gonzalez-Garcia, 
C. Pena-Garay, Phys. Rev. D {\bf 68}, 093003(2003);  J. Bahcall, M. C. 
Gonzalez-Garcia, C. Pena-Garay, hep-ph/0406294.
\bibitem{edm1}P. Nath, Phys. Rev. Lett. {\bf 66}, 2565(1991);
T. Ibrahim, P. Nath, Phys. Lett. {\bf B418}, 98; Phys. Rev.
{\bf D57}, 478(1998); {\bf 58}, 111301(1998); {\bf 61}, 093004(2000);
M. Brhlik, G. J. Good, G. L. Kane, {\it ibid.} {\bf 59}, 115004(1999);
A. Bartl, T. Gajdosik, W. Porod, P. Stockinger and H. Stremnitzer,
{\it ibid.} {\bf 60}, 073003(1999).
\bibitem{edm2}D. Chang, W. Keung and A. Pilaftsis, Phys. Rev. Lett.
{\bf 82}, 900(1999); A. Pilaftsis, Phys. Lett. {\bf B471}, 174(1999);
D. Chang, W. Chang and W. Keung, {\it ibid.} {\bf 478}, 239(2000).
\end{thebibliography}
\end{document}